\documentclass[12pt]{article}
\usepackage[latin1]{inputenc}

\usepackage{amsmath}
\usepackage{amsfonts}
\usepackage{amssymb}
\usepackage{graphicx}
\usepackage{geometry}
\usepackage{amssymb,epsfig}
\usepackage{hyperref}
\usepackage{subfig}

\pdfoutput=1

\makeatletter
\renewcommand\section{\@startsection {section}{1}{\z@}%
                                 {-3.5ex \@plus -1ex \@minus -.2ex}
                                   {2.3ex \@plus.2ex}%
                                   {\normalfont\large\bfseries}}
\renewcommand\subsection{\@startsection{subsection}{2}{\z@}%
                                   {-3.25ex\@plus -1ex \@minus -.2ex}%
                                     {1.5ex \@plus .2ex}%
                                     {\normalfont\bfseries}}
\renewcommand\subsubsection{\@startsection{subsubsection}{3}{\z@}%
                                   {-3.25ex\@plus -1ex \@minus -.2ex}%
                                     {1.5ex \@plus .2ex}%
                                     {\normalfont\itshape}}
\makeatother

\def\pplogo{\vbox{\kern-\headheight\kern -29pt
\halign{##&##\hfil\cr&{\ppnumber}\cr\rule{0pt}{2.5ex}&\ppdate\cr}}}
\makeatletter
\def\ps@firstpage{\ps@empty \def\@oddhead{\hss\pplogo}%
  \let\@evenhead\@oddhead 
}
\def\maketitle{\par
 \begingroup
 \def\thefootnote{\fnsymbol{footnote}}
 \def\@makefnmark{\hbox{$^{\@thefnmark}$\hss}}
 \if@twocolumn
 \twocolumn[\@maketitle]
 \else \newpage
 \global\@topnum\z@ \@maketitle \fi\thispagestyle{firstpage}\@thanks
 \endgroup
 \setcounter{footnote}{0}
 \let\maketitle\relax
 \let\@maketitle\relax
 \gdef\@thanks{}\gdef\@author{}\gdef\@title{}\let\thanks\relax}
\makeatother

\numberwithin{equation}{section}

\newcommand{\be}{\begin{equation}}
\newcommand{\bea}{\begin{eqnarray}}
\newcommand{\ee}{\end{equation}}
\newcommand{\eea}{\end{eqnarray}}
\newcommand\beq{\begin{equation}}
\newcommand\eeq{\end{equation}}

\renewcommand{\t}{\tilde}

\newcommand{\tu}{t_{UV}}

\begin{document}

\setcounter{page}0
\def\ppnumber{\vbox{\baselineskip14pt
}}
\def\ppdate{\footnotesize{SLAC-PUB-14552 ~~~ SU-ITP-11/44 ~~~ NSF-KITP-11-196}} \date{}

\author{Xi Dong,$^1$ Bart Horn,$^1$ Shunji Matsuura,$^{2,1}$ Eva Silverstein,$^1$ Gonzalo Torroba$^1$\\
[7mm]
{\normalsize \it $^1$Stanford Institute for Theoretical Physics }\\
{\normalsize  \it Department of Physics and SLAC, }\\
{\normalsize \it Stanford, CA 94305, USA}\\
{\normalsize \it $^2$Kavli Institute for Theoretical Physics }\\
{\normalsize  \it University of California, Santa Barbara, CA }\\
{\normalsize \it 93106, USA}\\
}

\bigskip
\title{\bf  FRW solutions and holography from uplifted AdS/CFT
\vskip 0.5cm}
\maketitle

\begin{abstract}

Starting from concrete AdS/CFT dual pairs, one can introduce ingredients which produce cosmological solutions, including metastable de Sitter
and its decay to non-accelerating FRW.
We present simple FRW solutions sourced by magnetic flavor branes and analyze correlation functions and particle and brane dynamics.  To obtain a holographic description,  we exhibit a time-dependent warped metric on the solution and interpret the resulting redshifted region as a Lorentzian low energy effective field theory in one fewer dimension.
At finite times, this theory has a finite cutoff, a propagating lower dimensional graviton and a finite covariant entropy bound, but at late times the lower dimensional Planck mass and entropy go off to infinity in a way that is dominated by contributions from the low energy effective theory. This opens up the possibility of a precise dual at late times.  We reproduce the time-dependent growth of the number of degrees of freedom in the system via a count of available microscopic states in the corresponding magnetic brane construction.

\end{abstract}
\bigskip
\newpage

\tableofcontents

\vskip 1cm

\section{Introduction:  keeping it real}\label{sec:intro}

At present we lack a complete theoretical framework for cosmology.  One approach to this problem is to try to organize cosmology holographically, building on the success of the AdS/CFT correspondence.  Doing so is not trivial for a number of reasons related to the tendency of cosmological solutions to mix with each other and the absence of a simple timelike boundary.  Dynamical gravity, or an integration over metrics, is part of the putative lower-dimensional dual in various attempts so far to generalize the AdS/CFT correspondence to cosmology and describe a complete set of observables; this includes dS/CFT~\cite{dSCFT}\ at least as it is interpreted in \cite{Maldacena:2002vr},\footnote{For this example, there may at least be a subset of observables which correspond to a precise non-gravitational CFT as described in \cite{Maldacena:2002vr}, where the CFT computes the wavefunction of the universe.  However, this wavefunction is a functional of the metric which one must ultimately integrate over.  See the recent work \cite{Anninos:2011jp}\ for more discussion of this question.} dS/dS~\cite{dSdS} and FRW/CFT~\cite{Freivogel:2006xu}.
Despite the lower dimensional gravity, the formulation of a significant part of the system in terms of a large matter sector is a nontrivial step, one which has recently been put on more solid footing microscopically~\cite{Dong:2010pm}.  Nonetheless it is important to understand whether a more precise formulation might exist.

The structure of UV complete cosmological solutions will likely be useful in answering this question.\footnote{In the somewhat analogous context of black hole physics,  study of concrete string theoretic examples led to microstate counts and ultimately the AdS/CFT correspondence.}
In this paper, we present and analyze concrete cosmological solutions which are sourced by a generic ingredient -- magnetic flavor branes -- used to uplift AdS/CFT systems \cite{Polchinski:2009ch}\ to cosmology.  With sufficiently many magnetic flavor branes, no nonsingular static solutions exist, but time-dependent solutions do exist which are nonsingular at late times; these solutions are nonsingular at all times if obtained from a bubble nucleation process.  (Another interesting class of dynamical F-theory solutions was studied in the earlier work~\cite{Kleban:2007kk}, which emphasized the point that no physical restriction on the number of 7-branes exists once the generic possibility of time dependence is included.)
We will introduce a holographic interpretation of this class of solutions, employing the following basic strategy.

First, we find a warped metric on our spacetime and interpret the two highly redshifted regions in terms of a pair of low energy effective theories.  This is a generalization of the observation in \cite{dSdS}\ that metastable $d$-dimensional de Sitter spacetime is a warped compactification with two throats and propagating $(d-1)$-dimensional gravity.\footnote{More recently, a description in terms of two CFTs coupled to gravity was motivated in another way by \cite{Harlow:2011ke}.}  This line of reasoning of course goes back to the original arguments \cite{Maldacena:1997re, OtherAdSCFT}\ that the highly redshifted core region of a stack of branes should be equivalent to a field theory, since it represents low energy degrees of freedom decoupled from the ambient Planck scale.  We verify that particles are stable in the infrared region, though color branes out on their approximate Coulomb branch propagate up the throat.  We call this phenomenon ``motion sickness''; as we will discuss later on it is not fatal.

The next step is to compute the $(d-1)$-dimensional Newton constant:\  this reveals that the $(d-1)$-dimensional graviton decouples at late times, in a way analogous to a Randall-Sundrum theory with the ``Planck brane" taken off to infinity.  This, and the growth of the entropy at late times~\cite{Harlow:2010my}, is consistent with the possibility of an ultimately precise holographic dual decoupled from gravity.  Although gravity decouples in this promising manner, we will see that the way the field theory induces a growing Planck mass is through a rapidly growing number of degrees of freedom, rather than via a growing cutoff on the effective field theory.  That is, the system at late times behaves like a theory which is holographic and non-gravitational, but with a finite cutoff for the dual theory.  Cutoff quantum field theory is in principle well defined, but many questions remain about its detailed implementation in gauge/gravity duality.\footnote{There has been interesting recent progress in relating radial slices to RG scale in AdS/CFT \cite{Heemskerk:2010hk}, but the detailed dictionary remains to be understood, and is subject to various important subtleties such as the fact that different types of gravity-side particles have different relationships between their energy and their radial position.}

In fact, we find a nontrivial match between the time-dependent number of degrees of freedom in the $(d-1)$-dimensional dual theory, computed using the gravity side in three different ways, with an estimate of the number of available microscopic degrees of freedom on the magnetic flavor branes responsible for the uplift to cosmology.   The states we count are drawn from the infinite algebras discussed in \cite{MITseven}, cut off at finite time by backreaction and topological consistency criteria.  As we will describe below, as currently formulated this count is consistent with basic group-theoretic requirements, but is not fully derived.  It is subject to two assumptions about unknown quantities -- the first is a plausible conjecture made but not proved in \cite{MITseven}, and the second regards the number of charged matter representations which arise.
With these assumptions, our count consistently reproduces the gravity-side result in a general class of solutions in different dimensions in a way that appears nontrivial, and generalizes the parametric microscopic estimate of the dS entropy of \cite{Dong:2010pm} to FRW cosmologies.
These results seem rather encouraging, and motivate further study of time-dependent field theories with sufficiently many magnetic flavors to provide candidate duals for cosmological solutions.

Our formulation of the holographic dual as a Lorentzian-signature field theory (or effective field theory) maintains standard reality and unitarity properties; in particular the number of degrees of freedom in the matter sector is a positive real number.
There are other interesting approaches to de Sitter or FRW holography which define the dual on a spacelike (Euclidean) surface, and it would be interesting to study the relation between these different formulations.\footnote{In particular, an interesting approach to a concrete example of dS/CFT can be found in \cite{DioTom}.}
It may be useful to note, however, that because of the ultimate requirement of integrating over metrics, the argument for defining the theory on the boundary of the spacetime does not trivially generalize from AdS to dS or FRW solutions. Microscopically, large-radius de Sitter solutions in string theory do not arise as a simple continuation of AdS solutions, which turns the flux imaginary in the Freund-Rubin solution.  The physically consistent metastable dS solutions that are known arise instead by uplifting AdS solutions with a more complicated collection of stress-energy sources. As we will see, defining a Lorentzian-signature dual via our warped metric does not a priori force us to forego a complete dual description:\ our warped solution decompactifies at late times, somewhat analogously to Randall-Sundrum with the Planck brane removed to infinity.

Another basic motivation for this work is to further develop our understanding of the structure of time-dependent and cosmological solutions in string theory.  We compute correlation functions of massive and massless particles in our geometry; the latter requires a careful treatment of pseudotachyon modes \cite{Aharony:2006ra}.
The structure of these correlation functions should tell us much more about holography on our solutions, the detailed analysis of which we leave for future work.  One intriguing feature is that the two-point function of Kaluza-Klein modes is a power law, rather than exponential.

This paper is organized as follows: In the next section we present FRW solutions sourced by magnetic flavor branes uplifting Freund-Rubin compactifications.  We exhibit a warped metric on the solution, indicating a low energy sector corresponding to an effective field theory.
In Section \ref{sec:dynamics}, we show that particles remain stably in the throat at late times, and color branes move up the throat.
This theory is cut off and coupled to gravity at finite times, but the Planck mass and the number of degrees of freedom go off to infinity at late times in a manner that is dominated by contributions of the warped region, raising the possibility that the dual completes to a precise non-gravitational theory in this limit.  We compute the number of field theoretic degrees of freedom in several macroscopic ways in Sections \ref{sec:FRWsol} and \ref{subsec:quasilocal}, and also present, in Section \ref{subsec:microcount}, a count of brane degrees of freedom which agrees with the macroscopic predictions given certain assumptions.
In Section \ref{sec:cf}, we study the two-point correlation functions of scalar fields in our solutions, in the massive and massless cases; a full derivation is relegated to Appendix \ref{sec:acf}.  Finally, we conclude in Section \ref{sec:fd}, and outline directions for future study.

\section{FRW solution sourced by magnetic flavor branes}\label{sec:FRWsol}

We would like to understand whether FRW cosmology in $d$ dimensions, which occurs for example after decays of metastable de Sitter, admits a $(d-1)$-dimensional holographic dual description.
Our strategy is to look for a warped metric on the FRW solutions derived from uplifted AdS/CFT solutions in string theory.  We then interpret the infrared region of the warped metric -- the region of strong gravitational redshift -- in terms of a dual effective field theory (EFT).  Finally, we analyze whether the EFT might become a self-contained quantum field theory (QFT) in the far future, since the entropy bound and the $(d-1)$-dimensional Planck mass go to infinity in that limit.

\subsection{Magnetic flavor branes}\label{subsec:magnetic-branes}

The simplest $AdS_d/CFT_{d-1}$ dual pairs arise from Freund-Rubin compactifications on a positively curved Einstein manifold $Y$ stabilized by flux.  These can be understood as the near-horizon backreacted solution obtained from color branes placed at the tip of a cone ${\cal C}$ with base $Y$.

We will uplift to cosmological solutions by adding heavy branes which reverse the sign of the curvature of $Y$.  Consider first the $AdS_5\times S^5$ solution of type IIB string theory, with the $S^5$ viewed as a Hopf fibration over a base $\mathbb{CP}^2$ (there are many similar examples with $S^5$ replaced by a more general Einstein space $Y$).  As discussed in~\cite{Polchinski:2009ch}, there is a natural ingredient which competes with the internal curvature:\ $(p,q)$ 7-branes at real codimension two on the $\mathbb{CP}^2$, wrapping the Hopf fiber circle and extended along $AdS_5$.  Such branes can be described using F-theory \cite{Ftheory}, which geometrizes the varying axio-dilaton, and one finds that 36 7-branes are required to exactly cancel the curvature of the $\mathbb{CP}^2$.  Similarly, 24 7-branes are required to exactly cancel the curvature of a $\mathbb{CP}^1$, which arises as the base of the Hopf fibration in examples with a compactification on $S^3$, such as $AdS_3\times S^3\times T^4$.  In the latter case, alternatively one can use ``stringy cosmic 5-branes''\cite{Greene:1989ya,Hellerman:2002ax}, elliptic fibrations with the torus fiber coming from the $T^4$. See for instance \cite{Dong:2010pm}, where SC5-branes together with other ingredients are used to cancel the curvature of $\mathbb{CP}^1.$

Let us denote the elliptic fibration over the base $B=\mathbb{CP}^m$ -- the $\mathbb{CP}^m$ with 7-branes at real codimension two -- by $\hat B$, and the entire uplifted compactification by $\hat Y$.  Parameterize the number $n$ of 7-branes or stringy cosmic branes in all cases by defining a quantity
\beq\label{Deltandef}
\Delta n \equiv n-n_\text{flat}
\eeq
such that $\Delta n=0$ corresponds to a flat uplifted base $\hat B$.

In the AdS case $\Delta n < 0$, such configurations including their backreaction on the geometry can be described relatively simply using F-theory.  On the field theory side, these systems have magnetic flavors, arising in the brane construction from $(p,q)$ strings stretching between D3-branes and the $(p,q)$ 7-branes \cite{Sen:1996vd,Banks:1996nj}.

Bringing 7-branes together in a time-independent manner generically introduces singularities.  For sufficiently few 7-branes, it is understood how these singularities are resolved physically, giving enhanced symmetries and/or light matter fields.
In a gauged linear sigma model (i.e.\ toric) description of the geometry of the elliptic fibration, singularities appear as additional branches in the target space~\cite{Witten:1993yc}.  A criterion for physically resolved singularities of these static solutions~\cite{Polchinski:2009ch} is that the central charge of the additional branch be less than that of the main target space of the sigma model.  In this case, one may formulate a brane construction with 7-branes intersecting at the tip of a cone, at which the color branes are placed.

This geometry and the backreacted solutions were described in~\cite{Aharony:1998xz, Polchinski:2009ch}. Its salient features are captured by the five dimensional theory obtained by compactifying on $S^5$ and adding the potential energy of the 7-branes. The effective potential in 5D Einstein frame is
\be\label{eq:Ueff}
{\cal U}\sim M_5^5 (R_f R^4)^{-2/3}
\left(\frac{R_f^2}{
R^4}+\frac{\Delta n}{
R^2}+\frac{N_c^2}{R^8 R_f^2}\right)
\eeq
where $R_f\sqrt{\alpha'}$ is the size of the fiber circle $S^1_f$, and $R\sqrt{\alpha'}$ is the size of the uplifted base ${\hat B}$.  The first term is from the metric flux of the $S^1$ fiber, and the second is the net contribution of the internal curvature and seven-branes.
In this F-theory setting there is generically no global mode of $g_s$.  There are additional scalar fields from 7-brane moduli, which are relatively flat as discussed in \cite{Polchinski:2009ch}.  A simple case to consider is one in which the dilaton is fixed at an $SL(2,{\mathbb Z})$ invariant point, via the mechanisms discussed in \cite{Aharony:1998xz}.
The third term comes from $N_c$ units of 5-form flux corresponding to the color branes.  The middle term will concern us most in this work; it comes from the 7-brane sources.

The 7-branes wrap $AdS_5 \times S^1_f$ times a two-cycle in the base, and in the dimensionally reduced theory we do not keep track of their positions in the compact directions.
As reviewed in \cite{Polchinski:2009ch}, the geometrical understanding \cite{Ftheory}\ of 7-branes as an elliptic fibration makes it possible to calculate their leading contribution to the curvature, and hence to the potential energy (\ref{eq:Ueff}). One can study the geometry by realizing it as the target space of a gauged linear sigma model \cite{Witten:1993yc}.  In this description, the beta function for the size of the negatively curved internal space has the same scaling but opposite sign as in the case of a $\mathbb{CP}^m$, by an amount $\Delta n$ that depends on the number of 7-branes.   The deformations of the 7-brane configuration are superpotential terms in the sigma model and are intrinsically lighter.
We can for convenience focus on configurations where the string coupling has been fixed at $g_s \sim 1$, enforced by appropriate combinations of 7-branes; it is also interesting to consider the orientifold limit \cite{Sen:1997gv}. The static solutions with $\Delta n<0$ are then described by minima of (\ref{eq:Ueff})~\cite{Polchinski:2009ch}.

Bringing $\Delta n\ge 0$ branes together in a {\it static} configuration leads to singularities which violate the above condition for allowed singularities, with the central charge of the singular branch being larger than that of the main target space. From the point of view of the description (\ref{eq:Ueff}), $\Delta n \ge 0$ leads to a decompactification limit. Moreover, in such a configuration the states that transform under the infinite algebras realized on $(p,q)$ 7-branes~\cite{MITseven}, which are broken for separated 7-branes, appear to come down to zero mass. These effects hint that an infinite set of degrees of freedom may ultimately be involved in formulating physics in the generic case of $\Delta n > 0$, and we will return to this point after developing a controlled gravitational description of uplifted solutions.

In general, we should allow for time-dependent backreacted solutions~\cite{Kleban:2007kk}.  As we will explain shortly, the 7-branes need not come together anywhere at late times, and in appropriate examples (such as \cite{Dong:2010pm}).  An initial singularity may be avoided by matching to a Coleman-de Luccia tunneling process in the past (as described in Appendix \ref{sec:acf}), though we will in any case focus on the late time physics in the present work.

\subsection{Solution and warping}\label{subsec:sol}

Let us now introduce our solutions for $\Delta n>0$.  It is interesting to analyze this class of solutions both from a ten-dimensional perspective and using the $d$-dimensional description obtained by compactification on the uplifted space $\hat Y$.
Below, we will exhibit a precise 10d solution, but let us begin with the $d$-dimensional description.

In the case $d=5$, we have an effective potential (\ref{eq:Ueff}) for the scalar fields $R$ and $R_f$.
With $\Delta n>0$, the 7-branes overcompensate the contribution of the curvature to the effective potential, so they turn the base $\mathbb{CP}^2$ into a net negatively curved space $\hat B$, whose curvature scales with $R$ as if it were a hyperbolic space.  All terms in the potential (\ref{eq:Ueff}) are positive in this case, and we look for time-dependent solutions where the radii evolve with time (along with the FRW scale factor $a(t)$ in the $d$-dimensional theory).

As we will see momentarily, the $d$-dimensional FRW equations along with the equations of motion for the scalar fields $R$ and $R_f$ admit a time-dependent solution where at late times $R$ and the scale factor expand with time, and the fiber circle $R_f$ remains constant.
The dominant term in the potential energy (\ref{eq:Ueff}) in this solution is the term proportional to $\Delta n/R^2$, since the others decay more quickly at large $R$ and fixed $R_f$.  In particular, the 5-form flux corresponding to the color branes is very subdominant at late times.  We will make contact with this in Section \ref{sec:dynamics}, when analyzing the dynamics of color branes in our solutions.

Specifically, we find the scale factor $a(t) \propto t$, while $R \propto t^{3/7}$ and $R_f$ approaches a constant.  The mass of a KK mode on the uplifted base is
\beq\label{massEframe}
m_{KK}\sim \frac{n_{KK}}{R}\times \frac{M_5}{(R^4 R_f)^{1/3}} \propto \frac{n_{KK}}{t}
\eeq
where the second factor here is the conversion to Einstein frame.
This means that, as in the original examples of AdS/CFT \cite{Maldacena:1997re}, there is no hierarchy between the internal dimensions and the curvature scale in $d$ dimensions.
It is likely possible to use the method developed in \cite{Polchinski:2009ch}\ to obtain a hierarchy of scales, but as we will see shortly our solution is very simple in 10 dimensions.

These scalings can be obtained self-consistently by noting that in this limit the dominant contribution to the energy is given by the 7-branes and curvature, while the fluxes dilute faster. The FRW equations become
\bea\label{eq:FRW}
4 \,\frac{\ddot a}{a}&=& -\frac{28}{3}\frac{\dot R^2}{R^2}+\frac{2}{9}M_5^2 (R^4 R_f)^{-2/3} \frac{\Delta n}{R^2}\nonumber\\
12 \,\frac{\dot a^2}{a^2}+ 12\,\frac{K}{a^2}&=&\frac{28}{3}\frac{\dot R^2}{R^2} +\frac{2}{3}M_5^2 (R^4 R_f)^{-2/3} \frac{\Delta n}{R^2}\,.
\eea
Here $K$ is the spatial curvature of the FRW metric. Looking for a solution of the form $a(t)=ct$, the first equation gives
\be\label{eq:Roft}
R(t) = \left(\frac{7M_5}{3 \sqrt{42}}\frac{\Delta n^{1/2}}{R_f^{1/3}} \right)^{3/7}\,t^{3/7}\,.
\ee
Plugging in the second equation yields $K=-1$ (i.e.\ an open FRW solution) and $c^2=7/3$ with $a(t)=c t$.  Note that there will also be a dynamical equation for $R_f(t)$;  however, analyzing this equation of motion shows that it is self-consistently frozen in place in the regime above.  This and other features of the solution will be very clear in the 10d solution we will present shortly.  Numerical studies of the equations of motion for $a(t),$ $R(t),$ and $R_f(t)$ show that the solution above is an attractor for a range of initial conditions.

Let us now analyze the 10d solution. So far we have focused on $(p,q)$ 7-branes, but similar considerations apply more generally to FRW cosmologies sourced by other branes realized as elliptic fibrations, such as stringy cosmic 5-branes. More generally, other types of sources may be involved.  For example, in decay from the metastable $dS_3$ solution in~\cite{Dong:2010pm}, the uplifting contribution arises in part from stringy cosmic 5-branes but also from other sources such as NS5-branes.  In general, it is interesting to consider the FRW phase corresponding to the leading source at late times.  The elliptic fibrations we consider here are natural sources which contribute to the curvature at leading order, and we will continue to focus on this case here.

We then consider a $d$-dimensional FRW spacetime and an internal space that can be described as a Hopf fibration over a base $\mathbb{CP}^m$. As we argued before, as far as the evolution of the size $R$ of the base goes, the uplifted base $\hat B$ (the elliptic fibration over $\mathbb{CP}^m$) behaves like a hyperbolic space $\mathbb{H}_{2m}$ of real dimension $2m$.
Compactifying on this, we find the following Ricci-flat string-frame metric, which is hence a vacuum solution of Einstein's equation:
\beq\label{csoln}
ds^2_{s} = -dt_s^2 + \frac{t_s^2}{c^2} dH_{d-1}^2 + \frac{t_s^2}{\hat c^2} d{\hat B}_{2m}^2 + dx_f^2,
\eeq
where $c^2=(d+2m-2)/(d-2)$, $\hat c^2=(d+2m-2)/(2m-1)$, and
\beq\label{hypmet}
dH_{d-1}^2=d\chi^2+\cosh^2\chi\, dH^2_{d-2}
\eeq
is the metric on a noncompact, unit hyperboloid of dimension $d-1$.  ${d{\hat B}^2_{2m}}$ is the metric on our uplifted, negatively curved compact base space of dimension $2m$.  Although in this work we are concerned with generic configurations with $\Delta n>0$ branes uplifting AdS/CFT solutions, the solution above also describes the dynamics of a compactification on $S^1\times \mathbb H_{2m}/\Gamma$, a circle times a compact hyperbolic space.  In that case, the dilaton is meaningful (as oppposed to in F-theory); nevertheless, since our solution is Ricci-flat in 10d string frame, the dilaton is not sourced in this solution.

In this solution we only included the effects of the flavor branes.  In the $\Delta n<0$ case of AdS/CFT, the flux corresponding to color branes plays a leading role in the backreacted solution.  However, in the present case at late times
the flux dilutes away and is subdominant, as we emphasized above. Furthermore, since the contribution from the metric flux (first term in (\ref{eq:Ueff})) can also be neglected and the fiber size becomes constant at late times, we have approximated the fiber direction by an $S^1$ factor in the geometry.\footnote{Although the metric flux is subdominant in the solution, it does affect the topology; in particular the fiber circle remains contractible.  This feature will play a role in our count of brane degrees of freedom in Section \ref{subsec:microcount}.  The nontrivial fibration of the circle is a feature of the AdS/CFT dual pair which we are uplifting to FRW cosmology, so 
the metric flux may be an important element even though its energetic contribution is subdominant.  More generally, it would be interesting to develop a holographic duality for the solution without any metric flux.}  In some cases, there may be additional transverse dimensions (such as the $T^4$ in models based on $AdS_3\times S^3\times T^4$), which we suppressed in the metric (\ref{csoln}).

Let us now compactify down to $d$ dimensions. The volume of the compactification manifold $\hat Y$ is
\begin{equation}
\text{Vol}(\hat Y)\propto R_fR^{2m}\propto t_s^{2m},
\end{equation}
where we have not kept track of time-independent coefficients and have used $R\sim t_s$ from \eqref{csoln}. Going to the $d$-dimensional Einstein frame
\begin{equation}
g_{\mu\nu,E}^{(d)}=\left(\frac{\text{Vol}(\hat Y) M_{10}^8}{M_d^{d-2}}\right)^{2/(d-2)}g_{\mu\nu,s}^{(d)},
\end{equation}
we get an FRW metric of the form
\beq\label{csolnEframe}
ds^2_{E}=-dt^2+c^2t^2 dH_{d-1}^2\;,
\ee
where
\beq\label{cformula}
c^2=\frac{d+2m-2}{d-2}\;,\qquad
t\propto t_s^{c^2}\,.
\eeq
Cases of particular interest are uplifts of $AdS_5\times S^5$, with $c^2=7/3$ ($d=5$, $m=2$) and uplifts of $AdS_3\times S^3\times T^4$, with $c^2=3$ ($d=3$, $m=1$). This reproduces the results obtained using the $d$-dimensional theory: $R(t) \propto t^{1/c^2}$, in agreement with \eqref{eq:Roft}.

Since $c>1$, the scale factor is expanding faster than in curvature dominated FRW (a.k.a.\ flat spacetime in Milne coordinates).  In general, $\Delta n>0$ corresponds to $c>1$, and we will find it very useful to contrast our results for $c>1$ with the case of flat space ($c=1$).  Our holographic interpretation will apply consistently for $c>1$, and will not apply to flat spacetime.

We may change variables by setting
\be\label{varchange}
t=(\eta^2-w^2)^{c/2}\;,\;\chi=\frac{1}{2}\,\log\frac{\eta+w}{\eta-w},
\ee
and the metric \eqref{csolnEframe} becomes
\beq\label{wetametric}
ds^2 = c^2 (\eta^2-w^2)^{c-1}\left(dw^2-d\eta^2+\eta^2 dH_{d-2}^2 \right).
\eeq
This metric exhibits warping for $c>1$, which corresponds to $\Delta n>0$.  We want to understand the spectrum and dynamics of degrees of freedom that are redshifted to low energies.

It is useful to consider a closely related time coordinate $t_{UV}=\eta^c$, giving metric
\beq\label{metrictuv}
ds^2 = c^2\left(t_{UV}^{2/c}-w^2\right)^{c-1} dw^2 + \left(1-\frac{w^2}{t_{UV}^{2/c}}\right)^{c-1}\left(-dt^2_{UV}+c^2 t^2_{UV}dH^2_{d-2}\right).
\eeq
On the  UV slice $w=0$, we have $t_{UV}=t$. In this metric the warp factor
\be\label{warpf}
f(w,\tu)\equiv \Big(1-\frac{w^2}{t_{UV}^{2/c}}\Big)^{(c-1)/2}
\ee
and the metric component $g_{ww}$ depend only weakly on the coordinate time at late times:
\beq\label{slowvar}
\left|\frac{\partial_{\tu} f}{\partial_w f}\right|\sim \left|\frac{\partial_{\tu} g_{ww}}{\partial_w g_{ww}}\right|\sim \tu^{-(1-1/c)} \to 0 \qquad {\rm as} \qquad \tu\to\infty
\eeq
\noindent where in the last equivalence we have evaluated a point at constant warp factor, i.e.\ constant $w/t^{1/c}_{UV}.$  This is not a covariant quantity, but neither is the redshifted energy and the small value of the ratio (\ref{slowvar}) may simplify some calculations at late times.

There are other ways of writing the FRW spacetime as a warped product metric: as a simple example, we may pass to the conformal time $T = \frac1c\log(t_{UV}/\ell)$ in the $(d-1)$-dimensional theory, where $\ell$ is an arbitrary length scale. We can absorb the scale factor $ct_{UV}$ (in $d-1$ dimensions) into the warp factor and write the warped metric as
\beq\label{constantslicing}
ds^2 = c^2(\ell^{2/c}e^{2T}-w^2)^{c-1}dw^2 + c^2\ell^2e^{2cT}\left(1-\frac{w^2}{\ell^{2/c}e^{2T}}\right)^{c-1}(-dT^2 + dH^2_{d-2}).
\eeq
The dual theory now lives on a static (non-expanding) space $\mathbb{R} \times \mathbb{H}_{d-2}.$  This is similar to the AdS/CFT correspondence written on global vs.\ Poincar\'{e} slicing, or other slicings with an expanding or static hyperboloid \cite{Horowitz:2009wm}.  These various slicings describe a dual theory living on different spacetimes.  In our case, a complication is the presence of time-dependent couplings and (in general) a time-dependent metric for the field theory. Furthermore, since the dual theory is not conformal, the conformal transformation that removes the factor $e^{2cT}$ also modifies the running couplings, as we discuss in Section \ref{subsec:parameterrelns}. Other, more general slicings may lead to even more complicated dual descriptions: for instance, using Gaussian normal coordinates starting from a central spatial slice (hyperbolic or otherwise) gives different $w$-dependences in the temporal and spatial warp factors.
Although we will stick to the simpler example given above in this work, it would be interesting to consider the existence of dual theories for more general slicings.

Again, using the warped metric given above, we wish to determine what light (meaning energy $\ll$ bulk Planck mass $M_d$) degrees
of freedom there are.  Given a region of strong gravitational redshift, i.e.\ a region of light states
in our gravity solutions, the system may have a right
to a field theory description as in the low energy regime of the usual AdS/CFT.

The proper time interval between two events of coordinate interval $\Delta\tu$ is
\beq\label{redshift}
\Delta T (w,\tu)=\Delta\tu \,\left(1-\frac{w^2}{\tu^{2/c}}\right)^{(c-1)/2}\,.
\eeq
This redshift factor is, of course, 1 for $c=1$ (flat spacetime), and for $c>1$ it is smaller than 1. This indicates gravitational redshift for probes of proper energy $\sim 1/\Delta T$ (fixed in units of the $d$-dimensional Planck mass $M_d$).  Energies of such probes are redshifted down by a factor of $f(w,\tu)$ defined in (\ref{warpf}).  As we mentioned above, this is time-dependent as well as dependent on the ``radial" scale $w$, but its $w$ dependence is stronger (\ref{slowvar}).  Slices of constant $w/{\tu}^{1/c}$ are then slices of constant scale.  In terms of the coordinates given in (\ref{varchange}), this corresponds simply to slices of constant $\chi$.

This effect arises in the absence of any flux, suggesting that the flavor branes (or more generally the geometry they source) support dynamical degrees of freedom in the EFT region.\footnote{In some examples of AdS/CFT obtained by color branes probing the tip of a cone, there are closed string moduli at the tip, for example ones corresponding to Fayet-Iliopoulos terms. The question was raised in those examples of whether these modes are dynamical. There, the fact that the cone itself was unwarped supports the conclusion that the FI term is a parameter, not a field. In more general cases such as ours, however, the answer may be different.}

It is important to note that basic degrees of freedom such as KK modes, oscillating closed strings, 7-7 strings and junctions, D-branes, and so on do not have fixed masses in units of $M_d$. Their masses depend on $t=(\tu^{2/c}-w^2)^{c/2}$, leading to $\tu^{2/c}-w^2$ dependence in $\Delta T$ in (\ref{redshift}).  This is analogous to radially-dependent masses in AdS/CFT.\footnote{For example, in AdS/CFT compactified on a circle, momentum modes on the circle become radially-dependent masses for which the redshift factor precisely cancels out, and there are many other examples one could consider.}  For KK modes on the uplifted base, KK and winding modes on the fiber, strings, and 7-7 strings/junctions we obtain respectively
\beq\label{masses}
m_{KK}\propto \frac{1}{t}\;,\quad
m_{f}\propto m_\text{str}\propto \frac{1}{t^{1-1/c^2}}\;,\quad
m_{77}\propto \frac{1}{t^{1-2/c^2}}\,.
\eeq
We will analyze their dynamics in Section \ref{sec:dynamics}.  For our examples, the specific values for the exponents are, respectively, $-1$, $-4/7$, $-1/7$ for $m=2$, $d=5$, and $-1$, $-2/3$, $-1/3$ for $m=1$, $d=3$.

\subsection{Planck mass in $d-1$ dimensions and its decoupling at late times}\label{subsec:PlanckM}

As in Randall-Sundrum theory (RS) \cite{Randall:1999vf}, we can compute the $(d-1)$-dimensional Newton constant $G_{N,d-1}$ by dimensionally reducing on the $w$ direction.  This yields
\begin{align}
\frac{1}{G_{N,d-1}}&\equiv M_{d-1}^{d-3}\sim M_d^{d-2}\int_0^{\tu^{1/c}} dw \sqrt{-\tilde{g}}\tilde{g}^{\tu\tu}\\
\label{Planckmass}
&\sim M_d^{d-2}\int_0^{\tu^{1/c}} dw \left(1-\frac{w^2}{\tu^{2/c}}\right)^{\frac{(d-2)(c-1)}{2}}\tu^{1-1/c}\sim M_d^{d-2}\tu.
\end{align}
\noindent where by $\tilde{g}_{\mu \nu}$ we mean the factors that appear in the $d$-dimensional metric but not in the $(d-1)$-dimensional metric.  Thus at finite times, we have a warped compactification with propagating $(d-1)$-dimensional gravity as in the dS/dS correspondence \cite{dSdS},
but as $\tu\to\infty$ gravity decouples.

This raises the possibility of a more precise field theory dual in the far future. A simple but nontrivial test of this possibility is the following.  In a general warped compactification \cite{Warped}, a diverging $(d-1)$-dimensional Planck mass can arise in (at least) two ways:
\smallskip

\noindent (1) A leading contribution may come from the warped throat (the effective field theory), as in RS with a Planck brane moving off to infinity.
In this case, as the Planck brane goes off to infinity the holographic dual becomes a pure QFT decoupled from gravity, i.e.\ the effective field theory completes to a full QFT.

\smallskip

\noindent (2) Instead, in a more general warped compactification the leading contribution may come from the volume of the compactification manifold, with the warped throat subdominant.  In this case, the effective field theory does not complete to a full QFT which captures the full system.
\smallskip

Let us check which of these possibilities is realized in our system.  First, let us elaborate on the behavior (1) above in the case of RS.  This consists of an $AdS_5$ spacetime
\beq\label{RSAdS}
ds^2 = \frac{r^2}{R_{AdS}^2}\left(-dt^2 + d\vec x^2\right) + \frac{R_{AdS}^2}{r^2}dr^2
\eeq
up to a finite radial scale $r_{UV}$.  The 4-dimensional Planck mass $M_4$ is given by dimensionally reducing on the radial direction:
\beq\label{RSMP}
M_4^2\sim M_5^3\int_0^{r_{UV}} dr \sqrt{-\tilde{g}}\tilde{g}^{tt}\sim M_5^3 \frac{r_{UV}^2}{R_{AdS}}\sim \tilde N_{dof, AdS}\Lambda_{c,RS}^2
\eeq
In the last relation here, which indicates that the Planck mass is induced by the field theory degrees of freedom, we used that the central charge of the field theory scales like $\tilde N_{dof, AdS}\sim M_5^3 R_{AdS}^3$ \cite{Susskind:1998dq}\ and that the energy scale of the cutoff is $\Lambda_{c,RS}=r_{UV}/R_{AdS}^2$.

For our purposes, it will be useful to belabor this result in the following way.  First, let us break up the calculation (\ref{RSMP}) into two pieces: the integral over $r$ from 0 to $\epsilon\,r_{UV}$, and the rest of the integral from $\epsilon\,r_{UV}$ to $r_{UV}$, where $\epsilon$ is a fixed constant between 0 and 1.  This separates an IR contribution $r<\epsilon\,r_{UV}$ (corresponding to energies below $\epsilon\,r_{UV}/R_{AdS}^2$) from a UV contribution for $r>\epsilon\,r_{UV}$, using an arbitrary reference scale $\epsilon\,r_{UV}/R_{AdS}^2$ that is fixed in terms of the UV cutoff as we increase $r_{UV}$.  The ratio of the two contributions is a constant as the Planck brane moves off to infinity.  In particular, the infrared region continues to contribute a leading piece to the 4-dimensional Planck mass.

Let us analyze the same question in our FRW case. First, define a scale $M_*$ dividing the UV and IR regions of our throat via
\beq\label{UVIRsplit}
\frac{M_*}{M_{UV}}\equiv \epsilon\; \Rightarrow \;w_*=\tu^{1/c}\sqrt{1-\epsilon^{2/(c-1)}}\,.
\eeq
In terms of this, we can work out the ratio of UV to IR contributions to the Planck mass (\ref{Planckmass}), obtaining
\beq\label{Planckratio}
\frac{UV}{IR}=\frac{\int_0^{\sqrt{1-\epsilon^{2/(c-1)}}}dy(1-y^2)^{(d-2)(c-1)/2} }{\int_{\sqrt{1-\epsilon^{2/(c-1)}}}^1 dy(1-y^2)^{(d-2)(c-1)/2} }=\text{constant}\,.
\eeq
Thus our setup behaves similarly to case (1), raising the possibility of a pure field theory dual capturing the FRW physics at late times.

\subsection{Covariant entropy bound}\label{subsec:bousso}

An important quantity that characterizes a field theory is its number of degrees of freedom.  For example, if we cut off a theory with a lattice, we require some number $\tilde N_\text{dof}$ of degrees of freedom per lattice point to define it.  That is, we denote the number of field theoretic degrees of freedom by $\tilde N_\text{dof}$.  We would like to understand this quantity in our putative holographic theory, generalizing the analysis given in \cite{Susskind:1998dq}.  There, an infrared cutoff on the radial coordinate in AdS was related to a UV cutoff in the corresponding QFT.
Even in ordinary AdS/CFT this UV cutoff is not understood very precisely, however:\ it is not literally a lattice cutoff since it does not break the isometries of the space on which the field theory lives, and the UV/IR relation works differently for different types of probes on the gravity side.  However, gravitational calculations of the entropy of thermal states and of the central charges in the field theory reproduce the behavior expected from an identification of
$\tilde N_\text{dof}$ with the number of degrees of freedom per lattice point, and we may revert to that language.

We will compute the time dependence of this quantity in several distinct ways in the present work, including a count of available states on the magnetic flavor branes, obtaining the same answer.
On the gravity side, a measure of the number of degrees of freedom is given by the covariant entropy bound~\cite{bousso} on the entropy passing through an observer's past light sheet.  As emphasized in \cite{Harlow:2010my}, for FRW solutions (unlike the metastable de Sitter phase) the entropy bound grows to infinity at late times.

Let us work this out explicitly in our solution.
We start by choosing a spherically symmetric set of coordinates on (\ref{csolnEframe})
\be\label{eq:slice1}
ds_E^2 = -dt^2 + c^2 t^2\left(\frac{dr^2}{1+r^2}+r^2 d\Omega_{d-2}^2 \right)\,.
\ee
Consider an observer at the origin ($r=0$) in our space at time $t_0$.  The past light cone of this observer is foliated by spheres of size $\rho=r(t)ct$, where $r(t)$ is determined by
\beq\label{rt}
\int_0^r \frac{dr}{\sqrt{1+r^2}}=-\int_{t_0}^t \frac{dt}{ct}\qquad
\Rightarrow\quad r(t)=\frac{1}{2}\left[\left(\frac{t_0}{t}\right)^{1/c}-\left(\frac{t}{t_0}\right)^{1/c}\right].
\eeq
For $c>1$, the sphere grows to a maximal size $\rho_{max}\propto t_0$ and then begins to shrink (because of the contraction of the FRW universe) as we go back in time.

The conjectured entropy bound \cite{bousso} is given by the area of this maximal sphere in Planck units.  From this we obtain, substituting the time $t_0$ of the observer by $t$,
\beq\label{eq:SAH}
{\cal S}\sim M_d^{d-2} t^{d-2}.
\eeq
(For $c=1$, the sphere never reaches a maximal size, but instead keeps growing, indicating a diverging entropy bound even at finite time.) In our case, the entropy going to infinity at late times also suggests the possibility of a precise dual of our FRW phase when $t \to \infty$; this jibes with the infinite warped throat we develop at late times in our solution.


\subsection{Basic relations among parameters}\label{subsec:parameterrelns}

We are now in a position to list some basic relations between several quantities in our system:  the $(d-1)$-dimensional Planck mass, the number of field theoretic degrees of freedom $\tilde N_\text{dof}$ in the effective field theory (EFT), and the cutoff $\Lambda_c$ of the EFT.  These relations will enable us to solve for their dependence on time $\tu$.  We will derive $\tilde N_{dof}$ independently using the quasilocal stress tensor below in \S\ref{subsec:quasilocal}, obtaining the same result for its dynamics as is predicted by the simple considerations of this section.

First, since the $(d-1)$-dimensional Planck mass is largely induced by the field theory (as we just found in the previous subsection), we have the relation
\beq\label{SPlanck}
\tilde N_\text{dof} \,\Lambda_c^{d-3}\sim\frac{1}{G_{N,d-1}}\equiv M_{d-1}^{d-3}\sim M_d^{d-2} t_{UV}
\eeq
where $\Lambda_c$ is the cutoff of our effective field theory.

In our $(d-1)$-dimensional theory, we expect a nontrivial quantum energy density.  If we assume
that this is an order one fraction of the source of Hubble expansion in the dual, we obtain
a second relation comes from the Friedmann equation in the $(d-1)$-dimensional theory:
\beq\label{Friedmann}
H_{d-1}^2 = \frac{1}{t_{UV}^2}\sim \tilde N_\text{dof}\Lambda_c^{d-1}G_{N,d-1}.
\eeq
Here $H_{d-1}=1/\tu$ is Hubble in the $(d-1)$-dimensional theory obtained by dimensional reduction on $w$ in (\ref{metrictuv}).

Putting these together, we find
\beq\label{tdep}
\tilde N_\text{dof}\sim M_d^{d-2}\tu^{d-2}\;,\qquad
\Lambda_c\sim\frac{1}{\tu}\,.
\eeq
This result is consistent with the entropy discussed in the last subsection if we assume the basic relation
\beq\label{Srelation}
{\cal S}\sim \tilde N_\text{dof} \Lambda_c^{d-2}\; \text{Vol}_{d-2}
\eeq
where $\text{Vol}_{d-2}\sim \tu^{d-2}$ is the volume of space in the dual theory.
The result (\ref{tdep}) is also consistent with the result for $\tilde N_\text{dof}$ below in (\ref{subsec:quasilocal}).

It is also possible to define the theory on a non-expanding lattice using the coordinatization \eqref{constantslicing}, where the $(d-1)$-dimensional theory lives on $\mathbb{R} \times \mathbb{H}_{d-2}.$  In this case a calculation analogous to \eqref{Planckmass} gives
\beq\label{SPlancknew}
\tilde N_\text{dof}\Lambda^{d-3}_c \sim \frac{1}{G_{N,d-1}} \sim M_d^{d-2} \ell e^{(d-2)cT} \sim \frac{M_d^{d-2}t^{d-2}_{UV}}{\ell^{d-3}}.
\eeq

In this case, the Friedmann equation requires that
\beq\label{Friedstatic}
0=H_{d-1}^2=G_{N,d-1}\rho +\frac{1}{\ell^2}
\eeq
where $\ell$ is the curvature scale of the hyperbolic spatial slices.  One implication of this is that the energy density must compensate the time dependence in the Newton constant.
If we assume again that the energy density is of order $\tilde N_\text{dof}\Lambda^{d-1}_c$, we obtain the relation
\beq\label{Friedmannnew}
G_{N,d-1} \tilde N_\text{dof} \Lambda^{d-1}_c \sim\frac{1}{\ell^2}
\eeq
In this case, we then get
\beq\label{tdepnew}
\tilde N_\text{dof} \sim M_d^{d-2}t^{d-2}_{UV}\;,\qquad
\Lambda_c \sim \text{constant}.
\eeq
Again, this agrees with the covariant entropy bound and with the independent derivation of $\tilde N_\text{dof}$ we find below in (\ref{subsec:quasilocal}).
Note that the cutoffs in (\ref{tdep}) and (\ref{tdepnew}) are related by a conformal rescaling:
to obtain the effective field theory on $\mathbb{R} \times \mathbb{H}_{d-2}$, we have to remove the overall factor $e^{2cT}$ in (\ref{constantslicing}) by
a conformal transformation.
This should be compared with the corresponding term in (\ref{metrictuv}) where there is no such factor.
Since our theory does not have a conformal symmetry, this conformal transformation changes the running coupling constants and other scale dependent quantities.
We can see this explicitly in that the gravitational coupling $G_{N,d-1}$  is different in the two cases.


For both slicings, the final result agrees with the scaling (\ref{eq:SAH}) from the covariant entropy bound, since at late times $t \sim t_{UV}$.
Thus if we think of the cutoff as a lattice cutoff, the system builds up entropy by accumulating degrees of freedom per lattice point, rather than by increasing the number of lattice points.  In Section \ref{subsec:microcount} we will provide an independent count of $\tilde N_\text{dof}$ using the magnetic brane construction, finding that the infinite store of degrees of freedom on our $\Delta n>0$ set of 7-branes, cut off by backreaction criteria, precisely reproduces this behavior.
As described above, these results are consistent with the possibility of a complete non-gravitational field theory dual at late times, albeit one with a finite cutoff for the field theory.  The growth of $\tilde N_\text{dof}$ is consistent with this interpretation:  in a field theory with time-dependent masses and couplings, the number of degrees of freedom below a fixed cutoff scale will generically change with time.  In our case, it increases rapidly.

\section{Dynamics of particles and branes}\label{sec:dynamics}

So far, we have presented our $d$-dimensional cosmological solution, exhibited a warped metric on it, and derived basic properties of its $(d-1)$-dimensional description, a candidate holographic dual. In this section we study the motion of particles and branes in our geometry. Their motion in the highly redshifted (infrared) region is related to the long distance dynamics of the putative holographic dual.
For simplicity we will consider the $d=5$ case; general dimensionalities $d$ can be studied in a similar fashion.

Our goals are twofold.  First, we would like to better understand the role of the color sector in our theory, given that the 5-form flux is subdominant in the solution and warping arises without it.  Secondly, we want to check whether the infrared degrees of freedom in the highly redshifted (warped) throat are stable.  In general, it would be interesting to understand what additional criteria -- beyond the presence of strong warping -- might need to be satisfied in order to obtain a holographic dual capturing the low energy degrees of freedom (see \cite{AdScrit}\ for some earlier discussion of this question).  One natural physical criterion is that the strength and variation of the warp factor be such that light particles remain in the warped region.\footnote{We thank D. Marolf and J. Polchinski for this suggestion.}

The results are as follows: color branes (D3-brane domain walls) are not stable in the IR region of our warped geometry, but particles (massive (\ref{masses}) and massless) do remain in the infrared region.  These facts can be seen in a simple way from the original metric (\ref{csoln}) (they follow equivalently from a similar analysis using the warped metric).

In the case $d=5$, $m=2$, the 10-dimensional string frame metric is (\ref{csoln})
\be\label{eq:10dstr}
ds^2 = -dt_s^2 + \frac{t_s^2}{c^2} (d\chi^2 + \cosh^2\chi\,{dH^2_3})+\frac{t_s^2}{c^2} {d{\hat B}^2_4} + dx_f^2\,,
\ee
where $dH_3^2$ is the metric on a unit 3-dimensional hyperbolic space $\mathbb H_3$ and $c^2=7/3$.
As mentioned above, slices of constant scale in our warped metric (\ref{metrictuv}) translate into slices of constant $\chi$, and the two infrared regions correspond to large $|\chi|$.

The isometries of the 4-dimensional hyperboloid, and the corresponding conserved momenta, imply that particles that start moving out in the $\chi$ direction continue propagating to larger $|\chi|$ as time goes on.
Massless particles head toward null infinity, traveling on null geodesics $d\chi=c\,dt_s/t_s$, so that
\beq\label{massless}
\chi_{\rm massless}(t_s)=c \log(t_s/t_0).
\eeq
Massive particles at a fixed point on $\mathbb H_3$ are governed by the Born-Infeld action
\beq\label{partBI}
S_{\rm massive}=-\int dt_s m(t_s)\sqrt{1-{\dot\chi}^2t_s^2/c^2}
\eeq
where the particle mass $m(t_s)$ can depend on time in our system.  In particular, the string-frame counterparts of (\ref{masses}) are:
\beq\label{massesstr}
m_{KK,s}\propto \frac{1}{t_s}\;,\quad
m_{str,s}\propto m_{f,s}\propto 1\;,\quad
m_{77,s}\propto t_s\;.
\eeq
The conserved momentum is
\beq\label{momentum}
p = \frac{\delta \mathcal{L}}{\delta \dot{\chi}} = \frac{m(t_s)\dot\chi t_s^2/c^2}{\sqrt{1-{\dot\chi}^2t_s^2/c^2}}.
\eeq
A particle at fixed $\chi=\chi_0$ (with $\dot\chi=0$ initially) stays at $\chi_0$; a particle moving toward the bottom of either throat continues to do so as $t_s\to\infty$.  We can solve for $\dot\chi$ using (\ref{momentum}), giving
\beq\label{dotchi}
\dot\chi =\frac{cp}{t_s\sqrt{p^2+m(t_s)^2t_s^2/c^2}}.
\eeq
From this we see that in all cases (\ref{massesstr}), $\dot\chi\to 0$ at late times.

At this point we should note that the two-dimensional slice of the spacetime traced out by $t$ and $\chi$ (at a fixed point on the $\mathbb H_{3}$) is simply flat space, with metric $-dt^2+c^2t^2 d\chi^2$.  In that subspace alone, the low energy region we are defining is a version of a Rindler horizon.
Since $c>1$ the spacetime is not flat overall, and the curvature affects generic particle trajectories and quantum mechanical wave packets, but a classical calculation of test particle trajectories does not sense this effect.

We get the same result, of course, by working directly in the $d$-dimensional warped metric given above.
There, to analyze particle dynamics we solve the equations of motion for a particle with a mass of the form $m(\eta,w)\sim (\eta^2-w^2)^\kappa$ (with $\kappa$ given by \eqref{masses}\ for some of the basic particles in our system).   In our solution, this is obtained by varying the action
\beq\label{BI}
S_{\rm massive}=-\int d\eta ~ m(\eta,w) ~ c(\eta^2-w^2)^{(c-1)/2}\sqrt{1-(dw/d\eta)^2}.
\eeq
At late $\eta,$ this yields a solution of the form
\beq
w=c_1\eta + \frac{c_2}{\eta^b}
\eeq
with $b = 2\kappa + (c-1)$.  In the examples discussed above (KK modes, closed strings, and 7-7 strings) $b \geq -1$, with equality occurring for the case of KK modes.  Again, this means that such particles stay in the infrared region we just defined if they start out there.  Note that when transformed back into regular coordinates $t, \chi,$ the subleading piece is important to allow the particles to propagate in time.

A D3-brane, on the other hand, experiences a force pushing it up the warped throat.  Its Born-Infeld action in 10d string frame is of the form
\beq\label{DthreeBI}
S_{D3}\sim -T_3\int dt_s\left(\frac {t_s}{c}\right)^3 \cosh^3\chi\,\sqrt{1-{\dot\chi}^2t_s^2/c^2},
\eeq
where $T_3$ is the D3-brane tension.  The $\cosh^3\chi$ factor introduces a force pushing the brane to smaller values of $|\chi|$.  It is straightforward to verify, as we will do momentarily, that as a result of this force the D3-branes come up the throat.  We will call this phenomenon ``motion sickness.''\footnote{This is a relative of Fermi seasickness, though the term ``Technicolor Yawn'' \cite{TY} most colorfully illustrates the connection between this class of phenomena and warped throats.}

Let us start from (\ref{eq:10dstr}) in the dimensionless conformal time $\tilde t=c \log(t_s/\ell)$, where $\ell$ is an arbitrary length scale.
The equation of motion that follows from the DBI action is
\be\label{eq:D3-eom}
\frac{d}{d \t t}\left(e^{4 \t t/c}\cosh^3\chi\,\frac{\chi'}{\sqrt{1-\chi'^2}} \right)+ 3 e^{4 \t t/c} \cosh^2\chi\,\sinh\chi\,\sqrt{1-\chi'^2}=0\,,
\ee
where we took the D3-brane position to depend on time only, and here $\chi'\equiv d\chi/d\tilde t$ represents the derivative with respect to the conformal time.

Deep in the IR of either of our warped throats (e.g.\ the one with $\chi \gg 1$), (\ref{eq:D3-eom}) can be reduced to a first order differential equation for $\chi'$.
Integrating this equation reveals that after some time ($\tilde t>1$) the probe D3-brane propagates up the throat and escapes from the IR region, reaching the UV slice $\chi = 0$ within a finite time.\footnote{If we set up the system with some 5-form flux (rather than starting with explicit D3-branes), it would be interesting to determine whether, and at what rate, D3-branes are nucleated.  The analysis in this section shows that once present, color branes do not remain in the warped region for all time.}
It is interesting to note that it takes
longer and longer for the branes to escape to this slice at later and later times $t_s$:  from (\ref{DthreeBI}), we have $|\dot\chi| < c/t_s$.

Let us remark briefly on the significance of the motion sickness and its time dependence, which should provide useful clues as to the nature of the dual theory.  Firstly, it is worth recalling that motion of color sector eigenvalues up the throat occurs in some familiar examples of AdS/CFT.  One example is the case of ${\cal N}=4$ supersymmetric Yang-Mills theory on a compact negatively curved space \cite{Horowitz:2009wm,negcurv}, where the eigenvalues are subject to a negative quadratic potential.  This theory is unitary, because the eigenvalues take forever to get to infinity.  The system is properly treated by putting the eigenvalues in a scattering state.  Another example is Fermi seasickness \cite{Hartnoll:2009ns}, where finite density effects draw color branes up the warped throat.  If one cuts off these theories by embedding them in a warped compactification then the effect would take a finite time as long as the Planck mass is finite.

A new element in the present case is that the magnetic branes support some warping by themselves, and the color sector is subdominant in the solution.
As we have just seen, particles stay down the warped throat created by the magnetic branes, suggesting that the holographic dual may be built from degrees of freedom living on them.

Altogether, there are two reasons motion sickness does not appear to be fatal in our system: (i) the fact that there is an infrared region in the absence of the color D3-branes, and (ii) the fact that even in familiar examples of gauge/gravity duality where the color branes are responsible for all the warping, unitarity is not necessarily sacrificed
in the presence of a potential which pushes color branes toward the UV.  There are two approaches one can take:  (1) eject the offending color branes (treating the color sector as negligible, since the flavor branes provide warping anyway), or (2) wait it out (keeping the color branes in the game, since the instability takes longer at larger $t_s$).

In particular, in trying to better understand the field theory dual it may remain useful to think about starting with a color sector in place, since the ejection of the branes takes longer and longer as $t_s\to\infty$.  As we will see in the next section, however, the number of degrees of freedom is well accounted for by junction states living on the 7-branes themselves.

\section{Degrees of freedom in FRW holography}\label{sec:Ndof}

In Section \ref{sec:FRWsol} we found that the covariant entropy bound and the number of degrees of freedom per lattice point in the holographic dual grow as $t^{d-2}$. Now we will suggest a microscopic explanation for this time-dependent growth in terms of states associated to 7-7 strings and string junctions. These are natural candidates to account for the growing $\tilde N_\text{dof}$, because bringing together $\Delta n>0$ $(p,q)$ 7-branes in a static way leads to  infinite dimensional algebras~\cite{MITseven} that are realized on light states. We make this counting and the assumptions which go into it precise in Section \ref{subsec:microcount}.

In the gravity side, the magnetic flavor branes lead to warping and to an IR region, as we discussed in the previous sections. In the AdS/CFT correspondence, a warped geometry produces a nontrivial quasilocal stress tensor that can be used to compute the CFT energy momentum tensor and the central charge. In Section \ref{subsec:quasilocal} we generalize this method to cosmological solutions. This provides an independent calculation of $\tilde N_\text{dof}$
that agrees with the microscopic count and with the results in \S\ref{subsec:parameterrelns}).

\subsection{A microscopic count of degrees of freedom in FRW holography}\label{subsec:microcount}

In this paper we are focusing on quantities we can calculate under control in our gravity solutions, determining from them various basic features of the putative cutoff field theory dual.  This includes several independent computations showing that the number of field theoretic degrees of freedom grows with time as $\tilde N_\text{dof}\propto \tu^{d-2}$.  In this section, we will seek a microscopic accounting of these states.  In general, such a count is not guaranteed to work in any simple way; even in systems with a weak coupling limit and a large number of unbroken supercharges, $\tilde N_\text{dof}$ generally suffers corrections in going from weak to strong coupling.  For example, the two are related by a factor of 3/4 in the ${\cal N}=4$ super Yang-Mills theory, and by a more nontrivial interpolation in other examples.  Nonetheless, it is interesting to ask whether any natural count of states reproduces the parametric behavior of $\tilde N_\text{dof}$ in a given example.

Rather than going to weak coupling, one may trade fluxes for branes, turning on the corresponding scalar fields, and study the microstates which are evident in that phase \cite{Silverstein:2003jp}.  For example, in the ${\cal N}=4$ super Yang-Mills theory if we go out on the Coulomb branch by introducing $N_c$ explicit D3-brane domain walls, the $N_c^2$ degrees of freedom of the dual gauge theory become more manifest.
In that example, of course, the D3-branes source the warping, accounting for the low energy sector that indicates the existence of the field theory dual \cite{Maldacena:1997re}.

In our solutions, the 7-branes source warping in themselves.  As with flavor branes in AdS/CFT, they do not
come together in the backreacted solution (our time dependent solution), but in the corresponding (singular) static
solution they intersect at a point (where the color branes are placed in the brane construction).
Since they source some warping (like color branes normally do), there is a possibility that there are fundamental
degrees of freedom of the dual theory that live on their intersection.  Since they dominate at late times
(and since the color branes suffer from motion sickness), they may account for the lion's share of $\tilde N_\text{dof}$.
We will explore this possibility in this section, finding rather encouraging results.

Consider string and string junction states stretching between 7-branes.
For the generic case of $\Delta n >0$, there is an infinite dimensional algebra generated by these junctions \cite{MITseven}.\footnote{There have been similar intriguing appearances of large algebras organizing states and/or dynamics in works such as \cite{otheralg}.}  If the fiber circle were not contractible, they could also a priori carry arbitrarily large momentum and winding quantum numbers around that direction.
However, they ultimately back react on the geometry, the fiber circle is ultimately contractible, and Kaluza-Klein momenta are cut off by the giant graviton effect \cite{McGreevy:2000cw}, so at any finite time only finitely many states are available.  In this section, we will estimate the total number of degrees of freedom $\tilde N_\text{dof}$ at late times by counting junctions up to a cutoff determined by backreaction and topology.  We will be concerned with the $t$-dependence of $\tilde N_\text{dof}$, and will not keep track of time-independent factors.\footnote{In the future it might be very interesting to analyze the factors arising from group theory to characterize the dual theory in more detail.  The present calculation suggests that this is on the right track.}  The resulting count of brane degrees of freedom will precisely match the behavior
\beq
\tilde N_\text{dof}\sim t_{UV}^{d-2}
\eeq
found above from macroscopic considerations (\ref{tdep}).  As we will describe below, this statement is based on two assumptions we will specify.

Let us parameterize a state by the number $n_\text{str}$ of strings it contains stretching among the 7-branes, the winding number $n_f$ on the fiber circle, and the momentum number $k_f$ on the fiber circle. We will ultimately analyze all cases, with various dimensionalities for the compact and noncompact directions.  Let us start with the specific case of $d=5$, $m=2$ (i.e.\ uplifted $AdS_5\times S^5$).

We can bound $n_\text{str}$ by the requirement that the core size of the set of $n_\text{str}$ strings not exceed $R$, to avoid strong backreaction.  The core size is determined from the gravitational potential,
which goes like $1/r^{d_\perp-2}$, where $d_\perp$ is the codimension (e.g.\ $1/r$ for particles in 3+1 dimensions, $1/r^4$ for D3-branes in ten dimensions, and so on).  In our case, we need to take into account that the fiber circle is much smaller than $R$, so the effective codimension of the strings is 7 rather than 8.
With $n_\text{str}$ strings, the core size is given by
\beq\label{gravpot}
\frac{n_\text{str}}{r_\text{core}^5}\sim 1.
\eeq
From this, the condition that this size not exceed the size $R$ of the base $\mathbb{CP}^2$ is
\beq\label{core}
r_\text{core}\sim n_\text{str}^{1/5} \le R \sim t_s \sim t^{3/7} \qquad \Rightarrow \quad n_\text{str}\le t^{15/7}.
\eeq

The strings can wind around the fiber circle also.  If they wind $R/R_f\sim t_s\sim t^{3/7}$ times, they can detect that the fiber is contractible.  So let us cut off the windings at
\beq\label{wind}
n_f\le t^{3/7}.
\eeq

The string may also have momentum $k_f/(R_f\sqrt{\alpha'})$ along the fiber circle.  There is not a topological cutoff on $k_f$ as for windings; the momentum is a conserved charge.  However, the tower of momentum modes does not go up forever.  Ultimately, it was shown in \cite{McGreevy:2000cw}\ that Kaluza-Klein momenta are naturally cut off in UV-complete string theoretic examples of AdS/CFT in exactly the right way to mirror the operator content of the dual field theory.  In the present case, there is an important crossover at a lower scale. We may view our states as bound states of Kaluza-Klein gravitons and $(p,q)$ string junctions.  For sufficiently small $k_f$, i.e.\ $k_f/R_f\ll R$, the energy of each stretched string is of order $R/\sqrt{\alpha'}$; the gravitons are well bound to the strings.  But when $k_f$ crosses over to become much larger than $RR_f$, they are not strongly bound and the system as a whole behaves like a set of relativistic Kaluza-Klein gravitons of total momentum $k_f/(R_f\sqrt{\alpha'})$. (Ultimately, as in \cite{McGreevy:2000cw}, these gravitons grow into giants.)  Since the gravitons are not strongly bound to the stretched strings (if bound at all), the state may no longer be fundamental. Because of this crossover, we will count $k_f$ only up to $R R_f\sim t^{3/7}$:
\beq\label{momenta}
k_f\le t^{3/7}.
\eeq

Finally, we need to determine whether there are additional combinatorial factors which depend on $n_\text{str}$ arising from the algebra generated by the junctions.  To start, let us consider the configurations classified in \cite{MITseven}.  There, one starts from a subset ${\cal S}_0$ of the $(p,q)$ 7-branes which generates a finite Lie algebra ${\cal G}_0$.  One then considers junctions with some prongs ending on the ${\cal S}_0$ branes, transforming under ${\cal G}_0$ according  to their weight vector $\lambda$ under ${\cal G}_0$.  The rest of the prongs of these junctions form a set of $n_Z$ strings with charge $(p,q)$; these end on the remaining 7-branes (or some subset of them) denoted collectively by $Z$.  The works \cite{MITseven}\ show that a junction with weight $\lambda$ and asymptotic charge $n_Z(p,q)$ satisfies a relation
\beq\label{MITrelation}
\lambda\cdot\lambda=-{\bf J}^2+n_Z^2(f(p,q)-1).
\eeq
Here ${\bf J}^2\ge -2$ is the self-intersection of the junction, and $f(p,q)$ is a specific function of the asymptotic charges  (see for example eqn.~(2.9) of the second reference in \cite{MITseven}).  Since $\lambda\cdot\lambda\ge 0$, $f(p,q)\ge 1$ in the generic cases where the full system builds up an infinite algebra ${\cal G}_{inf}$.  In particular, for $f(p,q)>1$ the right hand side of (\ref{MITrelation}) can become large and positive by increasing $n_Z$, and the equation may be satisfied with longer and longer weight vectors $\lambda$ in ${\cal G}_0$.  (This is in contrast to the cases with $f(p,q)<1$, where at most a finite number of weights can satisfy (\ref{MITrelation}).)
It is conjectured in \cite{MITseven}\ that states with ${\bf J}^2=-2$ have the special feature that they can be realized by ``Jordan Open Strings", strings that end on two of the 7-branes (crossing cuts emanating from the 7-branes in the process).  We will count these strings.

Given that, next we need to know if there are any multiplicities in the tower of available states which grow with $n_\text{str}$ at large $n_\text{str}$.
As mentioned below (5.52) of the first reference in \cite{MITseven}, given a weight vector of length squared $\lambda\cdot\lambda$, there is a degeneracy given by the dimension of its Weyl orbit.  However, this number cannot grow arbitrarily large; it is bounded by the dimension of the Weyl group of ${\cal G}_0$.  In particular, if we consider any individual representation of ${\cal G}_0$, the number of weight vectors degenerate with the longest weight vector is given by the Weyl group, and does not grow with $n$.

There are additional degeneracies in the weight lattice beyond those required by Weyl reflections, but
these involve multiple representations.  As long as the number of these representations which arise in our physical system do not grow too fast with $n_\text{str}$, the count we propose here holds up.  It would be interesting to analyze this explicitly for specific Lie groups.  This question of which charged matter representations appear in F theory compactifications has been studied previously in string theory for phenomenological model building, and similar techniques may be useful in the present case.

Next, let us consider more general configurations which build up $\Delta n > 0$.  One way to think about these is by iterating the procedure just described, considering junctions stretching between the entire set of 7-branes corresponding to the infinite group ${\cal G}_{inf}$, and an additional $(p,q)$ 7-brane.  This works similarly to the above case, except now the Cartan matrix of the ${\cal G}_{inf}$ set of sevenbranes is of indefinite signature.  This by itself would allow for an infinite number of weight vectors of a fixed length squared (schematically of the form $\lambda_+^2-\lambda_-^2$ given the indefinite metric).  So in the equation (\ref{MITrelation}) there would be an infinite number of solutions even for {\it finite} $n_Z$.  However, this infinite number is itself cut off by insisting that the corresponding junctions be made up of at most of order $n_\text{str}$ strings.  This limits $\lambda_+$ and $\lambda_-$ above to be at most of order $n_\text{str}$ in length, and counting their degeneracy is similar to the above count for finite groups.

Putting this all together, the number of available degrees of freedom is obtained from the product of the maximum values of $n_\text{str}$, $n_f$, and $k_f$:
\beq\label{Ndof5}
\tilde N_\text{dof}\sim(n_\text{str} n_f k_f)_\text{max} \sim t^{15/7+2\times 3/7} = t^3 = t^{d-2}
\eeq
for $d=5$.

Next, let us work out $\tilde N_\text{dof}$ for $d=3$, $m=1$ (i.e.\ an FRW uplift of $AdS_3\times S^3\times T^4$).  In this case, the $T^4$ is of fixed size, independent of $t$, so (\ref{gravpot}) becomes
\beq\label{gravpotII}
\frac{n_\text{str}}{r_\text{core}}\sim 1.
\eeq
As a result, (\ref{core}) becomes
\beq\label{coreII}
r\sim n_\text{str} \le R \sim t_s \sim t^{1/3} \qquad \Rightarrow \quad n_\text{str}\le t^{1/3}
\eeq
Again, the strings can wind around the fiber circle as well.  If they wind $R/R_f\sim t_s\sim t^{1/3}$ times, they see that the fiber is contractible.  So we cut off the windings at
\beq\label{windII}
n_f \le t^{1/3}
\eeq
and also the momentum
\beq\label{momentaII}
k_f\le t^{1/3}.
\eeq
Altogether we get
\beq\label{Ndof3}
\tilde N_\text{dof}\sim(n_\text{str} n_f k_f)_\text{max}\sim t = t^{d-2}
\eeq
for $d=3$.

Having checked it now for two examples, let us finally consider all cases, varying the dimensionalities
of the FRW spacetime and the uplifted internal base space.
For $d$ noncompact FRW dimensions and a $2m$ dimensional uplifted base space, together with one fiber circle and $9-d-2m$ toroidal directions, we have
\beq\label{generalbounds}
n_\text{str} \le t_s^{d+2m-4}, \qquad  n_f \le t_s, \qquad k_f\le t_s
\eeq
where $t_s$ is the 10-dimensional string frame time coordinate (\ref{csoln}).
So $\tilde N_\text{dof}$ goes like $t_s^{d+2m-2}$.  Recall that $t_s$ goes like
$t^{1/c^2}$ in terms of the Einstein frame time coordinate (\ref{csolnEframe}), and $c^2=(d+2m-2)/(d-2)$ (\ref{cformula}).
Therefore for the general case, we have
\beq
\tilde N_\text{dof} \sim t^{d-2}\,,
\eeq
whose time dependence agrees precisely with the macroscopic calculations of $\tilde N_\text{dof}$.

\subsection{Deriving $\tilde N_\text{dof}$ from the quasilocal stress tensor}\label{subsec:quasilocal}

In the AdS/CFT correspondence, there are several ways to compute the number of degrees of freedom per lattice point of the CFT. For instance, Brown and Henneaux \cite{Brown:1986nw} used the conformal structure of asymptotically AdS$_3$ to identify the central charge, which was rederived from different points of view in \cite{Banados:1994tn,Balasubramanian:1999re}. The conformal anomaly can also be computed from the variation of the renormalized effective action under a conformal transformation \cite{Witten:1998qj,Henningson:1998gx}. Those analyses, however, depend highly on the conformal symmetry, or on the asymptotic structure of AdS spacetime. Since our spacetime does not have such a structure, we need to apply other methods.

One possible method is to excite our FRW spacetime and produce a black hole with a hyperbolic horizon in the IR region. The horizon entropy will be identified with that of the effective field theory, from which we could extract $\tilde N_\text{dof}$. This is analogous to the entropy computation of an AdS black hole in \cite{Witten:1998zw}.

Another method, which will be applied here, is to compute Brown and York's quasilocal stress tensor \cite{Brown:1992br} and identify it with the expectation value of the stress tensor in the boundary theory. In the context of AdS/CFT, this was used to compute the Casimir energy and $\tilde N_\text{dof}$ of the boundary theory in \cite{Horowitz:1998ha,Balasubramanian:1999re}. This method can be generalized to the dS/dS correspondence \cite{dSdS} and to our FRW model. In order to see how it works, let us first review the definition of the quasilocal stress tensor and then use it to calculate $\tilde N_\text{dof}$ of the dual theories in both dS/dS and FRW.

\subsubsection{Quasilocal stress tensor}

Let us consider a spacetime manifold $\mathcal{M}$ with a timelike boundary $\partial \mathcal{M}$ and space-time metric $g_{\mu\nu}$. This boundary could be a regularized one at some coordinate cutoff $r_c$.  Let $n_{\mu}$ be the outward pointing normal vector to $\partial\mathcal M$, normalized so that $n_{\mu}n^{\mu}=1$. The induced metric on the boundary is given by
\begin{equation}
\gamma_{\mu\nu}=g_{\mu\nu}-n_{\mu}n_{\nu},
\end{equation}
where a pullback onto $\partial\mathcal M$ is understood. The Einstein action including a boundary term is
\begin{equation}
S=\frac{M_d^{d-2}}{2}\int _{\mathcal{M}}d^dx \sqrt{-g}(\mathcal R-2\Lambda)-M_d^{d-2}\int_{\partial \mathcal{M}}d^{d-1}x\sqrt{-\gamma}\Theta+M_d^{d-2}S_{\text{ct}}[\gamma_{ab}],
\end{equation}
where $\Theta$ is the trace of the extrinsic curvature of the boundary
\begin{equation}
\Theta_{\mu\nu}=-{\gamma_{\mu}}^{\rho}\nabla _{\rho}n_{\nu},
\end{equation}
$S_{\text{ct}}$ is a suitably chosen local functional of the intrinsic geometry $\gamma_{ab}$ of the boundary, and $a$, $b$ are coordinate indices on the boundary. The quasilocal stress tensor \cite{Brown:1992br} is given by
\begin{equation}
\tau^{ab}\equiv\frac{2}{\sqrt{-\gamma}}\frac{\delta S}{\delta \gamma_{ab}}=M_d^{d-2}\left(\Theta^{ab}-\Theta\gamma^{ab}+\frac{2}{\sqrt{-\gamma}}\frac{\delta S_{\text{ct}}}{\delta \gamma_{ab}}\right).
\end{equation}
The AdS/CFT correspondence relates the expectation value of the stress tensor $\langle T_{ab}\rangle$ in the dual field theory to the limit of the quasilocal stress tensor $\tau_{ab}$ as the regularized boundary at $r_c$ is taken to infinity:
\begin{equation}\label{Tcfres}
\sqrt{-h}h^{ab}\langle T_{bc}\rangle=\lim_{r_c\to\infty}\sqrt{-\gamma}\gamma^{ab}\tau_{bc},
\end{equation}
where $h_{ab}$, the background metric on which the dual field theory is defined, is related to the boundary metric $\gamma_{ab}$ by a conformal transformation. The counterterms in $S_{\text{ct}}[\gamma_{ab}]$ are chosen appropriately so as to cancel all divergences when the limit is taken. This has a natural interpretation on the dual field theory side:\ we use local counterterms to obtain a finite, renormalized expectation value of the stress tensor in the field theory.

\subsubsection{$\tilde N_\text{dof}$ in AdS/CFT and dS/dS}

Let us calculate the quasilocal stress tensor in AdS$_d$ and dS$_d$, both sliced by dS$_{d-1}$. The metric is given by
\begin{equation}
ds_{(A)dS_d}^2=R^2dw^2+f(w)^2ds_{dS_{d-1}}^2,\qquad
ds_{dS_{d-1}}^2=h_{ab}dx^adx^b,
\end{equation}
where $f(w)=\sinh w$ for AdS$_d$ and $\sin w$ for dS$_d$. Here $R$ is the curvature radius of both $ds_{(A)dS_d}^2$ and $ds_{dS_{d-1}}^2$.

In the AdS$_d$ case, the AdS/CFT correspondence says that the bulk is dual to a boundary conformal field theory living on dS$_{d-1}$. In the dS$_d$ case, the dS/dS correspondence \cite{dSdS} conjectures that the bulk is dual to two effective field theories living on dS$_{d-1}$, both of which are cut off at an energy scale $1/R$ and coupled to $(d-1)$-dimensional gravity and to each other.

Coming back to our calculation, let us choose a regularized boundary at $w=w_c$. It has an induced metric $\gamma_{ab}=f(w)^2h_{ab}$, where $h_{ab}$ is the dS$_{d-1}$ metric on which the field theory is defined. The extrinsic curvature tensor is
\begin{equation}
\Theta_{ab}=-\frac1Rf(w_c)f'(w_c)h_{ab},
\end{equation}
and before adding the counterterms the quasilocal stress tensor is given by
\begin{equation}\label{qstad}
\tau_{ab}=\frac{(d-2)M_d^{d-2}}{R}f(w_c)f'(w_c)h_{ab}=
\begin{cases}
\displaystyle\frac{(d-2)M_d^{d-2}}{R}(\sinh w_c\cosh w_c)h_{ab},&\text{for AdS}_d,\\
\displaystyle\frac{(d-2)M_d^{d-2}}{R}(\sin w_c\cos w_c)h_{ab},&\text{for dS}_d.
\end{cases}
\end{equation}
In the AdS$_d$ case, the stress tensor is divergent as $w_c$ goes to infinity, and the counterterms in $S_{\text{ct}}$ renormalize it to
\begin{equation}
\tau_{ab}\sim\frac{M_d^{d-2}}{R}e^{-(d-3)w_c}h_{ab}
\end{equation}
for large $w_c$, so that the limit in \eqref{Tcfres} becomes finite and gives $\langle T_{ab}\rangle\sim(M_d^{d-2}/R)h_{ab}$.\footnote{Strictly speaking, the limit is nonzero only for odd $d$. In even (bulk) dimensions, the limit vanishes and so does the trace anomaly of a CFT in odd (boundary) dimensions. We can still get $\tilde N_\text{dof}$ by other means.} Matching this to the expectation value of the CFT stress tensor on dS$_{d-1}$ with $\tilde N_\text{dof}$ degrees of freedom per lattice point
\begin{equation}
\langle T_{ab}\rangle\sim\frac{\tilde N_\text{dof}}{R^{d-1}}h_{ab},
\end{equation}
we arrive at the correct number of degrees of freedom per lattice point
\begin{equation}
\tilde N_\text{dof}\sim M_d^{d-2}R^{d-2}
\end{equation}
in the context of AdS/CFT. In fact we do not have to take $w_c$ to infinity to get this parametric result. We could set $w_c$ to be of order 1 and forget about the counterterms (because there are no divergences to cancel if $w_c$ is fixed at order 1, and including the counterterms would only change the numerical coefficients which we are not keeping track of). This gives the same parametric result for $\tilde N_\text{dof}$.

This perspective is important for the dS/dS correspondence, because in this case the $w$ coordinate is bounded and we cannot take $w_c$ to infinity. What we can do is fix $w_c$ at order 1, and by the same argument we get $\tilde N_\text{dof}\sim M_d^{d-2}R^{d-2}$. This means that the number of degrees of freedom per lattice point for the dual theory on dS$_{d-1}$ is parametrically the same as the Gibbons-Hawking entropy of dS$_d$. This was confirmed by the concrete brane construction for the dS/dS correspondence in \cite{Dong:2010pm}.

It is interesting to note that the quasilocal stress tensor \eqref{qstad} vanishes at $w_c=\pi/2$ in the dS$_d$ case. This does not contradict our previous result $\tilde N_\text{dof}\sim M_d^{d-2}R^{d-2}$, because one can reliably deduce $\tilde N_\text{dof}$ of the dual field theory only well below its cutoff. The ``UV slice'' $w_c=\pi/2$ corresponds exactly to the energy scale where the effective field theory is cut off and coupled to gravity. The vanishing of the quasilocal stress tensor is reminiscent of the gravitational dressing effect discussed in \cite{dSdS}. As we shall see in the next subsection, this also happens in our FRW spacetime.

Before going on, we should note that the advantage of being able to take $w_c$ to infinity in the AdS$_d$ case is that we can calculate the exact $\tilde N_\text{dof}$ including the numerical coefficients. To demonstrate this, let us work for simplicity in $d=3$. Only one counterterm is needed in this case, which is simply a cosmological constant on the boundary: $S_{\text{ct}}=-(1/R)\int_{\partial\mathcal M}\sqrt{-\gamma}$. The quasilocal stress tensor becomes
\begin{equation}
\tau_{ab}=\frac{M_3}{R}(\sinh w_c\cosh w_c-\sinh^2w_c)h_{ab},
\end{equation}
which gives $\langle T_{ab}\rangle=(M_3/2R)h_{ab}$ upon taking $w_c$ to infinity. The stress tensor from $\tilde N_\text{dof}$ free massless scalar fields in dS$_2$ may be calculated in a standard way,\footnote{For two-dimensional (and therefore conformally flat) spacetimes one can use Eq.~(6.136) of \cite{Birrell:1982ix}. Note that their metric convention is the opposite of ours.} giving $\langle T_{ab}\rangle=(\tilde N_\text{dof}/24\pi R^2)h_{ab}$. This can also be obtained from the trace anomaly $\langle T_a^a\rangle=(c/24\pi)\mathcal R$ of a two-dimensional CFT. Matching this result to the quasilocal stress tensor, we arrive at
\begin{equation}
c\equiv\tilde N_\text{dof}=12\pi M_3R,
\end{equation}
which agrees exactly with \cite{Henningson:1998gx,Balasubramanian:1999re}.

\subsubsection{$\tilde N_\text{dof}$ in the FRW dual}\label{subsubsec:nfrw}

Let us now calculate the quasilocal stress tensor in our FRW spacetime. We start with the warped metric
\begin{equation}
ds_d^2=c^2(\eta^2-w^2)^{c-1}(dw^2-d\eta^2+\eta^2dH_{d-2}^2),
\end{equation}
and choose the regularized boundary to be a surface of constant $\alpha$ where $\alpha$ is defined by $w=\alpha\eta$ with $0<\alpha<1$.
This is a hypersurface of constant energy scale relative to the UV cutoff: $\alpha\approx1$ corresponds to the IR, while $\alpha\ll1$ corresponds to the UV.
The exact value of $\alpha$ is not essential in our analysis, as long as it is of order 1. This is because our main interest is the parametric dependence of the quasilocal stress tensor on $\eta$ (or $t_{UV}$), and this is not affected by the exact location of the hypersurface. One may also consider RG flows in our FRW geometry, as in AdS or dS \cite{Alvarez:1998wr}--\cite{Sahakian:1999bd}, and relate $\tilde N_\text{dof}$ in the IR to its value in the UV. The $\eta$-dependence of $\tilde N_\text{dof}$ is not affected by the energy scale.\footnote{$\alpha=1$ is a singular surface and we do not set the hypersurface there. In a broader picture, this singular surface is replaced by the Coleman--de Luccia instanton and connected to the parent de Sitter space.}

From the coordinate transformation $\chi=\frac12\log\frac{\eta+w}{\eta-w}$, a hypersurface of constant $\alpha$ is also a hypersurface of constant $\chi=\text{arctanh}\,\alpha$, so it is much easier to calculate the quasilocal stress tensor in the usual FRW coordinates
\begin{equation}
ds_d^2=-dt^2+c^2t^2(d\chi^2+\cosh^2\chi\, dH_{d-2}^2).
\end{equation}
The induced metric on the regularized boundary at $\chi=\chi_c$ is given by
\begin{equation}
\gamma_{ab}dx^adx^b=-dt^2+c^2t^2\cosh^2\chi\, dH_{d-2}^2,
\end{equation}
and the extrinsic curvature can be calculated as
\begin{equation}
\Theta_{tt}=0,\qquad
\Theta_{ij}=-\frac{\tanh\chi_c}{ct}\gamma_{ij},
\end{equation}
where $i$, $j$ are the coordinate indices on $\mathbb H_{d-2}$. The quasilocal stress tensor is given by
\begin{equation}\label{qstds}
\tau_{tt}=(d-2)\frac{M_d^{d-2}}{ct}(\tanh\chi_c)\gamma_{tt},\qquad
\tau_{ij}=(d-3)\frac{M_d^{d-2}}{ct}(\tanh\chi_c)\gamma_{ij},
\end{equation}
It is interesting to note that this corresponds to a perfect fluid with equation of state $w=(d-3)/(d-2)$. It has vanishing pressure in $d=3$.

The next step is to calculate the expectation value of the stress tensor in the dual field theory, which lives on the background metric $h_{ab}=\gamma_{ab}$ and has $\tilde N_\text{dof}$ degrees of freedom per lattice point:
\begin{equation}
\langle T_{ab}\rangle\sim\frac{\tilde N_\text{dof}}{t^{d-1}}h_{ab}.
\end{equation}
This can be obtained either on dimensional grounds or by using Eq.~(6.134) of \cite{Birrell:1982ix}, noting that $\gamma_{ab}$ is conformally flat. Let us set $\chi_c$ to be of order 1 as this is equivalent to setting $\alpha$ to be of order 1. We match $\langle T_{ab}\rangle$ to the quasilocal stress tensor \eqref{qstds} and obtain
\begin{equation}
\tilde N_\text{dof}\sim M_d^{d-2}t_{UV}^{d-2},
\end{equation}
where we have replaced $t$ with $t_{UV}$ because on a hypersurface of constant $\alpha$ we have $t=(1-\alpha^2)^{c/2}t_{UV}\sim t_{UV}$ from \eqref{wetametric}. This is in agreement with the microscopic count of the number of degrees of freedom.

Just as in the dS/dS correspondence, the quasilocal stress tensor \eqref{qstds} vanishes at $\chi_c=0$, corresponding to the UV slice $w=0$. Again this suggests renormalization from the $(d-1)$-dimensional gravitational effects.

\section{Correlation functions}\label{sec:cf}

In this section we compute two-point functions for massive and massless scalar fields in our solution.  These should provide detailed information about the nature of the $(d-1)$-dimensional theory.  That theory lives on an FRW geometry, has time-dependent and scale-dependent couplings, and is necessarily strongly coupled in order to reproduce a large-radius gravity solution.  We will leave a complete interpretation of our Green's functions to future work, but will note some of their interesting features below.

\subsection{Massive Green's functions}\label{subsec:magf}

In AdS/CFT it is well understood how gravitational scattering amplitudes behave like field theory correlators.  For example, massive propagators in the bulk (which are exponentially suppressed in flat space) turn into power-law CFT two-point functions in the dual theory through the effects of the warp factor, which is a strong function of the radial distance $r$.  Because of the warp factor, geodesics do not go along fixed-$r$ trajectories; they can take advantage of the warp factor and go along a shorter path by moving in the radial direction.  This leads to the power law rather than exponential correlators.  (A simple discussion of this can be found in \cite{Susskind:1998dq}.)

Let us ask the analogous question in our case.  Our dual theory is formulated on an FRW spacetime and has a time and scale-dependent coupling corresponding to the radius $R$ of $\hat B$, which depends on $t=(\tu^{2/c}-w^2)^{c/2}$.  (As noted above (\ref{slowvar}), this quantity depends more strongly on the radial scale coordinate $w$ than on time $\tu$.)    As we saw in Section \ref{subsec:parameterrelns}\ and Section \ref{sec:Ndof}, our theory is cut off at a finite scale, but accumulates additional field theoretic degrees of freedom as time evolves forward.
As a result, the theory is not conformally invariant.  Nonetheless we will see that there is one sector of massive bulk fields -- KK modes on the base $\hat B$ -- which have power law correlators.

We will now compute the leading behavior of the 2-point Green's functions for particles of mass $m(t)\propto t^{-\alpha}$ in our geometry.  To find the dominant trajectory, we must solve the equations obtained by varying the action
\beq\label{BItwo}
S=-\int d\lambda\, m(t) \sqrt{\dot t^2-c^2t^2\left(\dot\chi^2+\cosh^2\chi\,\dot{\tilde\chi}^2+\dots\right)},
\eeq
where ``dot" represents a derivative with respect to the worldline coordinate $\lambda$ and $dH_{d-1}^2=d\chi^2+\cosh^2\chi (d \tilde \chi^2+ \sinh^2\t \chi\,d\Omega_{d-3}^2)$.  We are interested in computing the propagator between the points $(t, \chi, \t \chi)$ and $(t, \chi, \t \chi+ \Delta \t \chi)$.  Note that because of the time-dependent mass, the calculation of the dominant trajectory is not equivalent to a calculation of the geodesic distance between the two points in our spacetime.

Our result can only depend on time $t$ and on the geodesic distance $L$ between two points in $\mathbb H_{d-1}$ because of the isometries of $\mathbb H_{d-1}$.
For two points on the same $\chi$-surface, the geodesic distance $L$ in $\mathbb H_{d-1}$ is given by\footnote{This is the hyperbolic law of cosines, and it can be obtained by analytic continuation from a similar formula on a sphere, or simply from the dot product in the embedding space.}
\be\label{eq:gdistance}
\cosh L = \cosh^2 \chi\,\cosh\Delta \t \chi- \sinh^2 \chi\,.
\ee
The central (UV) slice of our warped geometry is at $\chi=0$.
We will be interested both in Green's functions formulated deep in the IR at $\chi\gg1$, and also in Green's functions formulated at the UV slice $\chi=0$.

Since the propagator can only depend on $t$ and on the geodesic distance $L$, it is actually easier to do the calculation at constant $\t \chi$ and varying $\chi$, and at the end replace this by $L$. Consider a particle with mass
\be
m(t) = \frac{n_\alpha}{t^\alpha}\,,
\ee
where $n_\alpha$ is a constant of dimension $1-\alpha$.  In the $d=5$ case, $\alpha=1$, $4/7$, and $1/7$ for KK modes, closed strings, and 7-7 strings, respectively. Define the new coordinates
\be\label{eq:change}
\tau = \frac{t^{1-\alpha}}{1-\alpha},\qquad
y= (1-\alpha)c\chi,\qquad
(\alpha\ne 1)
\ee
and
\beq\label{eq:changeKK}
\tau=\log(t/\ell),\qquad
y=c\chi,\qquad
(\alpha=1).
\eeq
Let us first address the KK modes. Setting $n_1\equiv n_{KK}$ we see that the action (\ref{BItwo}) at constant $\t \chi$ reduces to
\beq
S=-n_{KK}\int d\lambda\sqrt{\dot\tau^2-\dot y^2}\qquad
(\alpha=1).
\eeq
This is equivalent to the action for a particle with a constant mass in flat spacetime.  The resulting two-point function is $e^{-n_{KK}\Delta y}$, exponentially suppressed in $\Delta y=c\Delta\chi=cL$.

However, this is a {\it power law} suppression in the geodesic distance $\Delta X$ in the $(d-1)$-dimensional spacetime $-d\tu^2+c^2\tu^2dH_{d-2}^2$ on which the $d-1$ theory lives.  Working at $\chi=0$, we have the geodesic distance $L=\Delta\tilde\chi$ according to \eqref{eq:gdistance}. We can see the power law behavior by making a coordinate transformation
\be
X=\tu\sinh (c\tilde\chi)\;,\;T=\tu\cosh (c\tilde\chi)\,,
\ee
where the metric is Minkowskian on the plane spanned by $X$ and $T$.  Taking our two points at the same $T$, i.e.\ taking $\tilde\chi_1=-\tilde\chi_2=\Delta\tilde\chi/2$,  we obtain the geodesic distance (for $\Delta \tilde\chi \gg 1$)
\beq\label{DeltaX}
\Delta X =2t_{UV}\sinh(c\Delta\tilde\chi/2)\approx\tu e^{c\Delta\tilde\chi/2},
\eeq
and hence our exponential result in $\Delta\tilde\chi$ becomes a power law in $\Delta X$.  The Green's function is of the form
\beq\label{KKGreens}
G(\Delta x)\sim e^{-n_{KK}cL}\sim e^{-n_{KK}c\Delta\tilde\chi}\sim \frac{1}{(\Delta x)^{2n_{KK}}},
\eeq
where $\Delta x\equiv\Delta X/ct_{UV}$ is the comoving geodesic distance.

This result is quite intriguing, as we obtain power law correlators from a massive KK mode propagator, which is reminiscent of what happens in ordinary AdS/CFT. However, here the mechanism for this involves the time dependence of the mass.
The propagator deep in the IR (large $\chi$) and for large $\Delta \t \chi$, has an additional exponential suppression as a function of $\chi$,
\be\label{eq:GKK2}
G(\chi;\Delta x)\sim e^{-n_{KK}cL}\sim e^{-n_{KK}c(2\chi+\Delta\tilde\chi)}\sim e^{-2cn_{KK}\chi} \frac{1}{(\Delta x)^{2n_{KK}}}\,,
\ee
where we used $L \approx 2\chi + \Delta \t \chi$ from the large $\chi$, $\Delta\tilde\chi$ limit of \eqref{eq:gdistance}. Recalling that $\chi=\frac{1}{2}\log \frac{\eta+w}{\eta-w}$ (\ref{varchange}), we find that the propagator has a power law scaling with $w/\eta$. This energy scaling of the correlator is again characteristic of a CFT, where power-law wavefunction renormalization is produced by nontrivial anomalous dimensions.  Note that these formulas only apply in the $c>1$ case, since for $c=1$ the KK masses are constant, and we do not get warping or this power law behavior.

Next, let us analyze the correlation functions for other massive scalar fields, those for which $\alpha\ne 1$.
\be
S=-n_\alpha\,\int d\lambda\,\sqrt{\dot \tau^2 - \tau^2 \dot y^2}\qquad (\alpha\ne 1).
\ee
This action is equivalent to the action for a particle with constant mass $n_\alpha$ in the Milne universe, and the dominant trajectory is obtained by analytic continuation from flat space:
\be
S_{cl}= 2i n_\alpha \tau\,\sinh\,\frac{\Delta y}{2}\,,
\ee
where the sign can be determined by the $-i\epsilon$ prescription implicit in the square root. Approximating the propagator by $e^{i S_{cl}}$ and rewriting the action in terms of the geodesic distance $L$ obtains
\be\label{eq:massiveG}
G(t; L) \sim \exp \left[-2n_\alpha \frac{t^{1-\alpha}}{1-\alpha}\sinh\frac{(1-\alpha)cL}{2}\right]\,.
\ee

Let us consider the massive fields with $\alpha<1$ (closed strings and 7-7 strings) at large $\Delta\tilde\chi\gg 1$, in which case we have $L\approx\Delta\tilde\chi+2\log\cosh\chi$ from \eqref{eq:gdistance}. Note that this expression is valid for both small and large $\chi$.
In terms of the geodesic distance $\Delta X$ \eqref{DeltaX} in the $(d-1)$-dimensional FRW spacetime, the massive Green's function (\ref{eq:massiveG}) has (for large $\Delta\tilde\chi \gg 1$) an exponential suppression
\be
G(t; \Delta X) \sim\,\exp \left[-n_\alpha \frac{t^{1-\alpha}}{1-\alpha}\left(\frac{\Delta X}{t_{UV}}\right)^{1-\alpha}(\cosh\chi)^{c(1-\alpha)}\right]
\sim\exp\left[-n_\alpha \frac{(\Delta X)^{1-\alpha}}{1-\alpha}\right],
\ee
where we have used $t_{UV}=t(\cosh\chi)^c$ from the coordinate transformation \eqref{varchange}. It is interesting to note that the $\alpha<1$ Green's function in this form is independent of $\chi$ (although there could be $\chi$-dependent prefactors that we have not kept track of).

These $\alpha<1$ fields become parametrically heavy at late times, dying away at long distances as compared to the KK modes with their power law correlators.  But they remain part of the theory: as we have seen in Section \ref{subsec:microcount}, the set of 7-7 strings and junctions which do not strongly backreact on the geometry make a leading contribution to the count of degrees of freedom.
This is somewhat reminiscent of off-diagonal matrix elements in the ${\mathcal N}=4$ SYM theory out on its Coulomb branch:\ these are parametrically heavy at large AdS radius as compared to KK modes, but they cannot be decoupled from the system.

\subsection{Massless Green's functions}

One can potentially learn about the field theory dual of a warped gravity solution such as ours by studying the massless correlation functions.
In this section, we give the correlation function of a massless scalar field in our open FRW spacetime, considering the case where it comes from a Coleman--de Luccia (CdL) decay \cite{Coleman:1980aw}. Details on deriving correlation functions in a general $d$-dimensional CdL geometry, including an explicitly Lorentzian prescription without the need to analytically continue the eigenmode expansion, are presented in Appendix \ref{sec:acf}.

The calculation is of interest more generally, as it requires developing tools to treat pseudotachyon modes \cite{Aharony:2006ra}\ which arise from Hubble expansion.  These are infrared modes which do not oscillate with time, but also do not grow large enough to significantly affect the background solution.  Such modes arise for $c>1$, as explained in Section \ref{subsec:afrw}. In general, these may be treated by putting them in a scattering state and computing correlation functions as expectation values in this state.  The method we use here involves choosing a particular state, the Hartle-Hawking vacuum that one obtains from the analytic continuation of the Euclidean CdL geometry.

We consider a minimally coupled massless scalar field $\phi$ in our FRW background
\begin{equation}
ds_d^2=-dt^2+c^2t^2dH_{d-1}^2,\qquad
dH_{d-1}^2=d\chi^2+\cosh^2\chi\, dH_{d-2}^2,
\end{equation}
where the scale factor $a(t)=ct$ is correct at late times. If this FRW spacetime comes from a CdL decay, the scale factor has to behave like $a(t)=t+\mathcal O(t^3)$ for small $t$, so that the big bang $t=0$ is just a coordinate singularity.

We are interested in the large $t$ behavior of the equal-time correlator
\begin{equation}
G(t,\chi,\Delta\tilde\chi)\equiv\langle\phi(t,\chi,\Delta\tilde\chi)\phi(t,\chi,0)\rangle,
\end{equation}
where we have put the two points on a hypersurface of constant $\chi$, corresponding to a given energy scale as discussed in Section \ref{subsubsec:nfrw}. In particular, $\chi=0$ corresponds to the UV and $\chi\gg1$ corresponds to the IR of the dual field theory. We use $\Delta\tilde\chi$ to denote the geodesic separation of the two points in the $\mathbb H_{d-2}$ directions. In Appendix \ref{sec:acf} we organize the correlation function into an expansion in the large $t$ and large $\Delta\tilde\chi$ limit. The first two leading terms turn out to be
\begin{align}\label{Glead}
G_{\text{leading}}(t,\chi,\Delta\tilde\chi)&\sim\Delta\tilde\chi+2\log\cosh\chi,\\
\label{Gsublead}
G_{\text{subleading}}(t,\chi,\Delta\tilde\chi)&\sim\frac{t^{(d-2)(\sqrt{1-1/c^2}-1)}}{\cosh^{d-2}\chi}e^{-(d-2)\Delta\tilde\chi/2}(\Delta\tilde\chi+2\log\cosh\chi),
\end{align}
up to terms that are ``pure gauge''.\footnote{This is because the massless correlator on a compact space such as the Euclidean CdL geometry is only well-defined up to arbitrary constants and linear functions of the coordinates. See a more detailed discussion in Section \ref{subsec:aeuc}.} The same leading term was obtained in \cite{Freivogel:2006xu}.

We rewrite the correlation function in terms of the geodesic distance $\Delta X$ in $d-1$ dimensions, which was derived in Section \ref{subsec:magf} to be $\Delta X\approx t_{UV}e^{c\Delta\tilde\chi/2}$. From the coordinate transformation \eqref{varchange}, $t_{UV}$ and $t$ are related by
\begin{equation}\label{tuv}
t_{UV}=t(\cosh\chi)^c.
\end{equation}
The leading contribution \eqref{Glead} to the correlation function becomes
\begin{equation}
G_{\text{leading}}(t,\chi,\Delta X)\sim\log\frac{\Delta X}{t_{UV}}+c\log\cosh\chi=\log\frac{\Delta X}{t},
\end{equation}
where we have used \eqref{tuv} in the last equality. We note that this leading term is present for both $c>1$ and $c=1$, and it is independent of the dimension $d$. This suggests an interpretation of it as the contribution from the zero mode of $\phi$ localized on the UV slice.

The subleading contribution to the correlation function in terms of $\Delta X$ is
\begin{align}\label{Gsubleading}
G_{\text{subleading}}(t,\chi,\Delta X)&\sim\frac{t^{(d-2)(\sqrt{1-1/c^2}-1)}}{\cosh^{d-2}\chi}\left(\frac{t_{UV}}{\Delta X}\right)^{(d-2)/c}\left(\log\frac{\Delta X}{t_{UV}}+c\log\cosh\chi\right)\\
&=\frac{t^{(d-2)(\sqrt{1-1/c^2}-1+1/c)}}{(\Delta X)^{(d-2)/c}}\log\frac{\Delta X}{t},
\end{align}
where we have used \eqref{tuv} in the second line. This term is dominated by contributions from the pseudotachyonic modes, and is present only for $c>1$. This is the case we have focused on in this paper, which has a warped geometry with a consistent description in terms of a low-energy $(d-1)$-dimensional dual.

In order to holographically interpret these results for $c>1$ in detail, it is necessary to understand better the behavior of a strongly coupled field theory with time and scale dependent couplings on a $(d-1)$-dimensional FRW background.  The strong coupling will limit our ability to compute, but it will be interesting to see if one can determine enough about this theory to make detailed comparisons with results such as \eqref{Gsubleading}.  We will leave an in-depth study of this for future work.

\section{Future directions:  Magnetic flavors and time-dependent QFT}\label{sec:fd}

This work raises many interesting questions, which we have only begun to explore.  In this section we describe a few of them.

On the gravity side, we have seen that in the presence of sufficiently many magnetic flavor branes to uplift AdS/CFT to cosmology, one has simple time-dependent solutions, whereas the would-be static solutions are singular (as observed before in \cite{Kleban:2007kk}\ via another class of time-dependent F-theory solutions).  We have exhibited a warped metric on these solutions, and have obtained results consistent with the interpretation of the corresponding low energy region as a $(d-1)$-dimensional holographic dual.

It would be very interesting to understand directly from a field-theoretic point of view where the distinction between $\Delta n<0$ and $\Delta n\ge 0$ comes from.  If there is a condition on the number of magnetic flavors (holding fixed other quantities) which corresponds to $\Delta n<0$,
the above results may suggest that time-dependent effects in field theory could change this condition.  Time dependent couplings affect the scaling dimensions of the corresponding operators in the effective Lagrangian.\footnote{One interesting new effect of time-dependent couplings in quantum field theory was discussed in \cite{Bao:2011pa}.}  On sufficiently short timescales, of course, the system becomes insensitive to the time-dependent couplings.  Perhaps this is related to the finite cutoff we are left with in our late-time non-gravitational field theories.

Even at finite times, the holographic description of our system will be very interesting to develop further.
The dual theory has time-dependent running couplings and (depending on our choice of conformal frame) in general lives on its own FRW geometry.  These features render it somewhat complicated to match in detail the correlation functions computed in Section \ref{sec:cf} and Appendix \ref{sec:acf} to quantities in the holographic dual, a task we therefore leave to (near-)future work.

A crucial step in a holographic formulation of cosmology is the identification of the correct microscopic degrees of freedom. In the solutions presented in this work, such states are predominantly given by string junctions extended between the magnetic flavor branes. We computed the number of degrees of freedom $\tilde N_\text{dof}$ in several ways, and found agreement between the gravity and field theory sides. This generalizes the microscopic calculation of the de Sitter entropy of \cite{Dong:2010pm} to a time-dependent cosmology. While some of our results are specific to the class of solutions presented in Section \ref{sec:FRWsol}, our methods suggest a concrete framework for holographic cosmology, which may have wider applications. Yet another method
would be to put the system at finite temperature by adding a black brane and computing its entropy -- we hope to pursue this calculation further.

Finally, we expect our results here to translate into a clearer understanding of appropriate observables in cosmological spacetimes.  There have been a number of interesting attempts to find a consistent framework in which to define probabilities in cosmology.  In our view, this program will likely benefit from concrete study of the structure of UV complete time-dependent backgrounds, in the same way as occurred in black hole physics.  There, analyzing the dynamics of brane solutions in string theory led to black hole entropy counts and ultimately to the AdS/CFT correspondence.  The details of particular examples may not be of central importance in the end, but can provide much needed checks and may lead us toward the right principles.

On this note, it would be extremely interesting to determine to what extent the magnetic flavor branes which play a key role here (and in other landscape constructions which use F-theory \cite{Douglas:2006es}) are generic in cosmological solutions with holographic duals.  In general there is a longer list of ``uplifting" possibilities, including starting from a larger total dimensionality $D>10$ (the more generic case) and/or compactifying on more general curved manifolds \cite{Silverstein:2004id}.  Magnetic flavors bring in a new source of strong coupling -- in addition to the strong 't Hooft coupling required to formulate large-radius gravity -- and it would be interesting to further understand their role in the theory.

\section*{Acknowledgments}
We are grateful to D. Anninos, D. Harlow, S. Kachru, J. Kaplan, J. Maldacena, D. Marolf, D. Park, J. Polchinski, N. Seiberg, Y. Sekino, S. Shenker, D. Stanford, and L. Susskind for useful discussions. We thank the Perimeter Institute for Theoretical Physics, Centro de Ciencias de Benasque Pedro Pascual,
the Institute for Advanced Study and Instituto Balseiro for hospitality during various phases of this work.
SM acknowledges the hospitality of the Stanford Institute for Theoretical Physics while this work was in progress.
This research was supported in part by the National Science Foundation under grant PHY05-51164, by NSF grant PHY-0244728, and by the DOE under contract DE-AC03-76SF00515.  BH is also supported by a William K. Bowes Jr.\ Stanford Graduate Fellowship.

\appendix

\section{Correlation functions in general CdL geometry}\label{sec:acf}
In this appendix we develop an exact formalism for calculating correlation functions in a general CdL geometry. One of our main goals is an explicitly Lorentzian formula \eqref{gfinal} for the real-time correlator in the FRW region of the CdL geometry, so that one can apply it without having to analytically continue the Euclidean modes into Lorentzian signature. This is very desirable in any calculation away from the thin-wall limit, because in general the exact eigenmodes are not analytically tractable. One can certainly make approximations or compute them numerically, but these are not useful for analytic continuation because one cannot reliably continue an approximate solution.\footnote{A toy example is the $\tanh x$ function, which is well approximated by 1 for large, real $x$, but we certainly cannot continue this approximate result to the imaginary axis. A class of exact CdL solutions showing this explicitly will appear in \cite{Dong:Harlow}.} Instead, we can derive an exact Lorentzian formula \eqref{gfinal}, and if necessary make approximations from there. This is the approach of this appendix.

In Section \ref{subsec:aeuc} we deveop a Euclidean prescription \eqref{gexp2} for the correlation function. This is a relatively straightforward generalization of the discussion in Appendix~A of \cite{Freivogel:2006xu} to arbitrary dimensions and arbitrary scale factors.\footnote{Correlation functions in general dimensions were also calculated in \cite{Park:2008ki} in the thin-wall limit. In this appendix we work more generally, partly because our FRW spacetime with $c>1$ is not in the thin-wall limit.}  In Section \ref{subsec:alor} we analytically continue \eqref{gexp2} to an exact, Lorentzian formula \eqref{gfinal}, with extra care given to the choice of integration contours. Section \ref{subsec:afrw} applies the Lorentzian prescription to our FRW spacetime. Finally, in Section \ref{subsec:amass} we briefly outline how to generalize this to a massive correlation function.

\subsection{Euclidean prescription}\label{subsec:aeuc}

The Euclidean CdL instanton in $d$ dimensions is characterized by the isometries of a $(d-1)$-dimensional sphere. It is the Euclidean version of a closed FRW universe. Let us work in the conformal coordinates:
\begin{equation}
ds_d^2=a(X)^2(dX^2+d\Omega_{d-1}^2),\qquad
d\Omega_{d-1}^2=d\theta^2+\sin^2\theta\,d\Omega_{d-2}^2,
\end{equation}
where $X$ varies from $-\infty$ to $+\infty$. The smoothness of the instanton requires that the scale factor vanish as $\xi+\mathcal O(\xi^3)$ at the two tips, where $\xi$ is the proper coordinate in the $X$ direction. In the conformal coordinates this means $a(X)$ approaches (up to constant factors) $e^X$ as $X\to-\infty$, and approaches $e^{-X}$ as $X\to+\infty$.

We calculate the two-point function of a minimally coupled massless scalar field $\phi$. It is convenient to consider the rescaled field
\begin{equation}
\hat\phi=a(X)^{k_0}\phi,\qquad
\text{where}\quad k_0\equiv\frac{d-2}{2},
\end{equation}
which has a canonical kinetic term. Note that $k_0$ is the momentum gap for the normalizable eigenmodes of the Laplacian on the unit $(d-1)$-dimensional hyperboloid.  We let $G$ and $\hat G$ denote the correlator of $\phi$ and $\hat\phi$ respectively. The rotational symmetry of the $d-1$ sphere may be used to bring one of the two points to $\theta=0$, so that the correlator $\hat G(X,X',\theta)$ depends on no other angular coordinates and satisfies the Laplace equation with a delta function source:
\begin{equation}
\left[-\partial_X^2-\nabla_{S^{d-1}}^2+U(X)\right]\hat G(X,X',\theta)=\frac{\delta(X-X')\delta(\theta)}{S_{d-2}\sin^{d-2}\theta},
\end{equation}
where $S_n=2\pi^{(n+1)/2}/\Gamma((n+1)/2)$ is the volume of the unit $n$-sphere, and the potential $U(X)$ is defined as
\begin{equation}
U(X)=\frac{b''(X)}{b(X)}=k_0^2-\frac{k_0\Phi'^2+d\,a(X)^2V(\Phi)}{2(d-1)},\qquad
\text{where}\quad b(X)\equiv a(X)^{k_0}.
\end{equation}
Here $\Phi$ is the scalar field sourcing the metric and has a canonical kinetic term. Its potential\footnote
{It is important not to confuse the ``Schr\"odinger potential'' $U(X)$ with the potential $V(\Phi)$ which sources the geometry. It is also important to distinguish the two scalar fields $\phi$ and $\Phi$.}
$V(\Phi)$ between the two tunneling points $\Phi(X=\pm\infty)$ is nonnegative for decays from de Sitter to de Sitter or to FRW with a zero cosmological constant (if one assumes the null energy condition), so $U(X)$ is bounded from above by $k_0^2$ and asymptotes to this bound as $X\to\pm\infty$.

The correlator $\hat G(X,X',\theta)$ can be obtained as an expansion in the eigenmodes of the ``Schr\"odinger operator'' $[-\partial_X^2+U(X)]$. The result is
\begin{equation}\label{gexp}
\hat G(X,X',\theta)=\int_{-\infty}^{+\infty}\frac{dk}{2\pi}u_k(X)u_k^*(X')G_k(\theta)+\sum_\kappa u_{i\kappa}(X)u_{i\kappa}^*(X')G_{i\kappa}(\theta),
\end{equation}
where the first term is an integral over the orthonormal continuum modes satisfying the ``Schr\"odinger equation"
\begin{equation}\label{schr}
\left[-\partial_X^2+U(X)\right]u_k(X)=(k^2+k_0^2)u_k(X),
\end{equation}
with the boundary conditions
\begin{gather}
u_k(X)\to e^{ikX}+R(k)e^{-ikX}\quad(X\to-\infty),\quad
u_k(X)\to T(k)e^{ikX}\quad(X\to+\infty),\\
u_{-k}(X)\to T_r(k)e^{-ikX}\quad(X\to-\infty),\quad
u_{-k}(X)\to e^{-ikX}+R_r(k)e^{ikX}\quad(X\to+\infty),
\end{gather}
where $k>0$ is understood. One can show that these coefficients are related by
\begin{equation}
T(k)=T_r(k),\qquad
\frac{R(k)}{R_r^*(k)}=-\frac{T(k)}{T_r^*(k)},\qquad
|R(k)|^2+|T(k)|^2=1.
\end{equation}

The second term in \eqref{gexp} is a discrete sum over the normalized bound states. They satisfy the Schr\"odinger equation \eqref{schr} with $k=i\kappa$. There is always at least one bound state for a compact Euclidean CdL: the zero mode $u_{ik_0}(X)\propto a(X)^{k_0}$. For $d\le4$ one can show that this is the only bound state using a technique of supersymmetric quantum mechanics \cite{Freivogel:2006xu}.
For $d>4$ there may be additional bound states.

The $G_k(\theta)$ that appears in \eqref{gexp} is the Green's function on $S^{d-1}$ with an appropriate mass:
\begin{equation}\label{eq:Gk}
\left[-\nabla_{S^{d-1}}^2+(k^2+k_0^2)\right]G_k(\theta)=\frac{\delta(\theta)}{S_{d-2}\sin^{d-2}\theta}.
\end{equation}
The solution may be written in terms of the hypergeometric function:
\begin{equation}\label{gk}
G_k(\theta)=\frac{(2\sqrt\pi)^{1-d}}{\Gamma\left(\frac{d-1}{2}\right)}\Gamma(k_0+ik)\Gamma(k_0-ik)\;{}_2F_1\left(k_0+ik,k_0-ik,\frac{d-1}{2},\cos^2\frac{\theta}{2}\right),
\end{equation}
which is a meromorphic function in $k$. It has simple poles located at
\begin{equation}
k=\pm i(k_0+n),\qquad
n=0,1,2,\cdots
\end{equation}
due to the Gamma functions in \eqref{gk}. In even dimensions the hypergeometric function simplifies to an elementary function. For example, in $d=4$ \eqref{gk} as a function of complex $\theta$ becomes
\begin{equation}
G_k(\theta)=\frac{\sinh\left[k(\pi-\theta)\right]}{4\pi\sinh(k\pi)\sin\theta},\qquad
\text{for}\quad0<\text{Re}(\theta)<2\pi.
\end{equation}

The Green's function $G_{i\kappa}(\theta)$ for the bound states is also given by \eqref{gk}. The only subtlety appears when we consider the zero energy bound state $a(X)^{k_0}$ with $k=ik_0$, for which $G_k(\theta)$ hits a pole and diverges. This reflects the fact that the massless correlator on a compact space is ill-defined because we cannot put a single source there without violating Gauss's law.\footnote{The left hand side of (\ref{eq:Gk}) integrates to zero, but the right hand side does not. This may be solved by subtracting an inhomogeneous term proportional to the inverse of the volume of the compact space.} The two-point function for the derivative of a massless field is well-defined and physical, so one may say that the massless correlator is well-defined up to arbitrary constant and/or linear terms in the coordinates. These terms are called ``pure gauge''. Therefore we may get a finite massless Green's function by subtracting an infinite constant from $G_{ik_0}(\theta)$. For example in $d=4$ one prescription is
\begin{equation}
G_{ik_0}(\theta)=\lim_{k\to ik_0}\left[G_k(\theta)-\frac{1}{2\pi^2(k^2+1)}\right]+\frac{1}{8\pi^2}
=\frac{\cot\theta}{4\pi}\left(1-\frac\theta\pi\right).
\end{equation}

\begin{figure}[!htbp]
\centering
\includegraphics[width=0.5\textwidth]{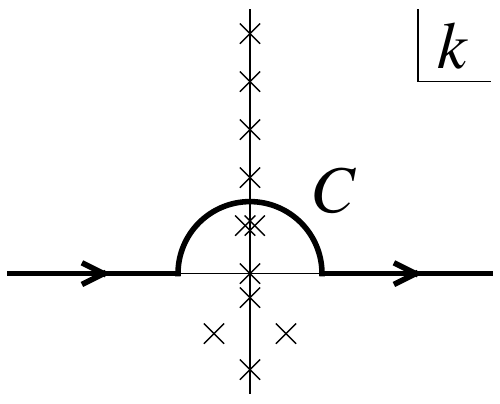}
\caption{Contour $C$ going from $k=-\infty$ above the double pole $k=ik_0$ to $k=+\infty$. Remember $k_0\equiv(d-2)/2$. The simple poles above the double pole are located at $k=i(k_0+n)$, $n=1,2,3,\cdots$. The locations of the simple poles below the double pole are for illustration purposes only and should not be taken too seriously. They depend on the reflection coefficient $R(k)$, and also on whether we are considering the second integrand in \eqref{gexp2} or in \eqref{gfinal}. The pair of poles away from the imaginary axis are resonance poles.}
\label{contour}
\end{figure}

\begin{figure}[!htbp]
\centering
\subfloat{\label{contoura}\includegraphics[width=0.44\textwidth]{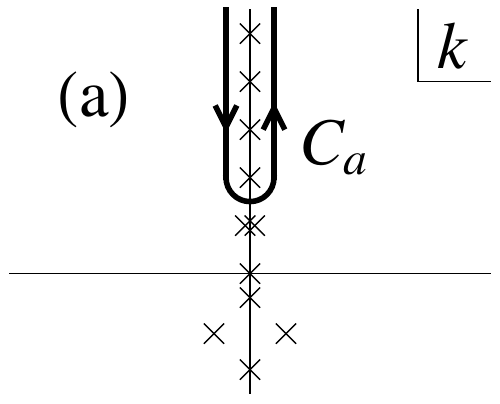}}
\qquad\qquad
\subfloat{\label{contourb}\includegraphics[width=0.44\textwidth]{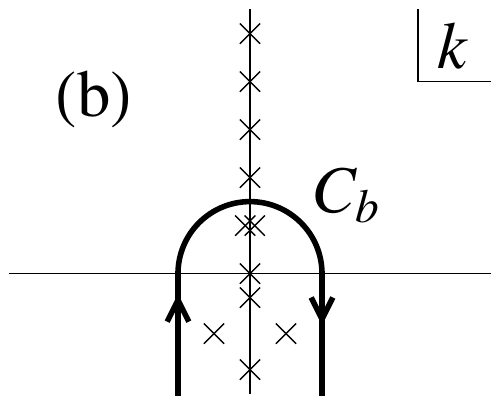}}
\caption{(a) Contour $C_a$ surrounding simple poles at $k=i(k_0+n)$, $n=1,2,3,\cdots$. (b) Coutour $C_b$ surrounding a double pole at $k=ik_0$ and simple poles below $ik_0$. In general there could also be resonance poles away from the imaginary axis in the lower half plane, in which case the contour $C_b$ needs to enclose them as well.}
\end{figure}

Before we analytically continue to the Lorentzian signature, let us rewrite the expansion \eqref{gexp} in a simpler form. Instead of using $u_k(X)$ we might want to write the correlator in terms of eigenmodes that have simpler asymptotic behavior as $X\to-\infty$, as this is where we cross to the FRW region. Let us call this new set of eigenmodes $v_k(X)$, defined with the boundary condition $v_k(X)\to e^{ikX}$ as $X\to-\infty$ for both positive and negative $k$. We may write them in terms of the old eigenmodes:
\begin{equation}
v_{k}(X)=\frac{u_{-k}^*(X)}{T_r^*(k)},\qquad
v_{-k}(X)=\frac{u_{-k}(X)}{T_r(k)},\qquad
k>0.
\end{equation}
After some algebra, one finds that in terms of the new eigenmodes the continuum contribution in \eqref{gexp} becomes
\begin{equation}\label{gcexp}
\hat G_c(X,X',\theta)=\int_{-\infty}^{+\infty}\frac{dk}{2\pi}\left[v_k(X)v_{-k}(X')+R(k)v_{-k}(X)v_{-k}(X')\right]G_k(\theta),
\end{equation}
where the reflection coefficient $R(k)$ is extended from $k>0$ to a meromorphic function on the complex $k$-plane. For $k<0$ one can show $R(k)=R^*(-k)$. One may ask what we have achieved by rewriting \eqref{gexp} as \eqref{gcexp}. It turns out\footnote{This may be shown by expanding the Green's function in terms of eigenmodes of the Laplacian on $S^{d-1}$ (discarding the zero mode). This expansion exactly matches the sum of contributions from the simple poles enclosed by $C_a$ as shown in Figure \ref{contoura}.} that the bound-state contribution in \eqref{gexp} can be accounted for, up to pure gauge, by simply deforming the integration contour for the second term in \eqref{gcexp}. Specifically, the massless correlator \eqref{gexp} simplifies to
\begin{equation}\label{gexp2}
\hat G(X,X',\theta)=\int_{-\infty}^{+\infty}\frac{dk}{2\pi}v_k(X)v_{-k}(X')G_k(\theta)+\int_{C}\frac{dk}{2\pi}R(k)v_{-k}(X)v_{-k}(X')G_k(\theta),
\end{equation}
where $C$ is a contour that goes from $k=-\infty$ above the double pole $k=ik_0$ to $k=+\infty$, as shown in Figure \ref{contour}.
In the limit $X$, $X'\to-\infty$, the second integrand in \eqref{gexp2} asymptotes to $e^{-ik(X+X')}$, so we may push the contour $C$ up to $C_a$, picking up the simple poles at $k=i(k_0+n)$, $n=1,2,3,\cdots$. This is shown in Figure \ref{contoura}.

Note that $R(k)$ has a simple pole at $k=ik_0$ (corresponding to the zero mode) and a zero at $k=-ik_0$, so the second integrand in \eqref{gexp2} has a double pole at $ik_0$ and is regular at $-ik_0$. It is precisely the residue at the double pole that cancels with the zero energy bound state contribution up to pure gauge. For $d>4$ there may be additional bound states, corresponding to additional simple poles between $k=0$ and $k=ik_0$. There could also be resonance poles, all of which must lie in the lower half plane. As we will see in the next two subsections, these additional poles need to be taken into account when calculating the exact correlation function, but at the order we work, they do not contribute to our final result.

\subsection{Lorentzian prescription}\label{subsec:alor}

Now that we have \eqref{gexp2} for the massless correlator in the Euclidean CdL geometry, we can analytically continue it to the FRW geometry
\begin{equation}
ds_d^2=a(T)^2(-dT^2+dH_{d-1}^2),\qquad
dH_{d-1}^2=dr^2+\sinh^2r\,d\Omega_{d-2}^2.
\end{equation}
The prescription for the analytic continuation in our conformal coordinates is given by
\begin{equation}
X\to T+\frac{i\pi}{2},\quad
\theta\to ir,\quad
a(X)\to ia(T),\quad
v_k(X)\to v_k(T)e^{-k\pi/2},
\end{equation}
so that $v_k(T)$ similarly satisfies a ``Schr\"odinger equation''
\begin{equation}\label{schrl}
\left[-\partial_T^2+U(T)\right]v_k(T)=(k^2+k_0^2)v_k(T),\quad
U(T)\equiv\frac{\ddot b(T)}{b(T)},\quad
b(T)\equiv a(T)^{k_0},
\end{equation}
with the simple boundary condition $v_k(T)\to e^{ikT}$ as $T\to-\infty$.

The Lorentzian correlator $\hat G(T,T',r)$ satisfies
\begin{equation}
\left[-\partial_T^2+\nabla_{\mathbb H_{d-1}}^2+U(T)\right]\hat G(T,T',r)=\frac{i\delta(T-T')\delta(r)}{S_{d-2}\sinh^{d-2}r},
\end{equation}
and is therefore given (for sufficiently negative $T+T'$) by analytically continuing the Euclidean correlator \eqref{gexp2}:
\begin{align}
\hat G(T,T',r)&=i^d\hat G(X\to T+\frac{i\pi}{2},X'\to T'+\frac{i\pi}{2},\theta\to ir)
\equiv\hat G_1+\hat G_2\\
\label{gexpl}
&=\int_{-\infty}^{+\infty}\frac{dk}{2\pi}v_k(T)v_{-k}(T')G_k(r)+\int_{C_a}\frac{dk}{2\pi}R(k)v_{-k}(T)v_{-k}(T')\tilde G_k(r),
\end{align}
where $G_k(r)$ and $\tilde G_k(r)$ are defined as
\begin{equation}\label{ggtilde}
G_k(r)=i^dG_k(\theta\to ir),\qquad
\tilde G_k(r)=i^de^{k\pi}G_k(\theta\to ir)
\end{equation}
except for the subtlety mentioned below for $\tilde G_k(r)$. Let us call the two integrals in \eqref{gexpl} $\hat G_1$ and $\hat G_2$ respectively. Note that our choice of the contour $C_a$ in $\hat G_2$ is correct only for sufficiently negative\footnote{One can see from \eqref{gtasym} that ``sufficiently negative'' here means $T+T'\ll -r$.}
$T+T'$. On the other hand, we are most interested in the massless correlator at late times. If we na\"{i}vely increase $T+T'$ without deforming the contour $C_a$, at some point the integral would diverge. A standard way of solving this problem is to pull the contour down to $C$ which goes from $k=-\infty$ above the double pole $k=ik_0$ to $k=+\infty$, as shown in Figure \ref{contour}. Once this is done and the new integral on $C$ agrees with the old one on $C_a$ for sufficiently negative $T+T'$, we simply use the new integral as the definition for any values of $T+T'$. To evaluate this new integral for large $T+T'$, one can legally deform the contour from $C$ to $C_b$ as shown in Figure \ref{contourb} (at least for some of the terms in the integrand), picking up the double and simple poles below the contour. This last step will be done on a case-by-case basis in the next subsection. Here we first obtain the correct expression for the new integral on $C$ that is valid for any $T$ and $T'$.

It turns out that we cannot na\"{i}vely deform the contour from $C_a$ to $C$ in $\hat G_2$. If we do that, the integral on $C$ would diverge, as we will show shortly. This is different from the situation before the analytic continuation, where it does not matter whether we choose $C$ or $C_a$, as long as $X+X'$ is sufficiently negative. The main difference is the additional $e^{k\pi}$ factor inside $\tilde G_k(r)$, which comes from analytically continuing $v_{-k}(X)v_{-k}(X')$ to $v_{-k}(T)v_{-k}(T')$ in $\hat G_2$.

A hint that something subtle is going on is that we could multiply the integrand of $\hat G_2$ by any number of $e^{2(k-ik_0)\pi}$ factors without changing the integral on $C_a$. The reason is simply that the integral on $C_a$ is equal to the sum of residues at the simple poles $k=i(k_0+n)$, $n=1,2,3,\cdots$, and $e^{2(k-ik_0)\pi}$ is simply 1 at all these poles, so none of the residues are changed. On the other hand, factors such as $e^{2(k-ik_0)\pi}$ definitely matter when the integral is performed on $C$. In fact, there is a unique prescription for this factor for which the integral on $C$ is actually convergent. This is the correct prescription, which we will find in exact form below.

Let us use the connection formula for the hypergeometric function
\begin{multline}
{}_2F_1(a,b,c,z)=\frac{\Gamma(c)\Gamma(b-a)}{\Gamma(b)\Gamma(c-a)}(-z)^{-a}{}_2F_1\left(a,1+a-c,1+a-b,\frac1z\right)\\
+\frac{\Gamma(c)\Gamma(a-b)}{\Gamma(a)\Gamma(c-b)}(-z)^{-b}{}_2F_1\left(b,1+b-c,1+b-a,\frac1z\right)
\end{multline}
to decompose the $\tilde G_k(r)$ into two terms:
\begin{align}
\tilde G_k(r)&=i^de^{k\pi}\frac{(2\sqrt\pi)^{1-d}}{\Gamma\left(\frac{d-1}{2}\right)}\Gamma(k_0+ik)\Gamma(k_0-ik)\;{}_2F_1\left(k_0+ik,k_0-ik,\frac{d-1}{2},\cosh^2\frac r2\right)\\
\nonumber
&=-\frac{1}{4\pi^{d/2}(4z)^{k_0}}\left[e^{2k\pi}(4z)^{-ik}\Gamma(k_0+ik)\Gamma(-ik)\;{}_2F_1\left(k_0+ik,\frac12+ik,1+2ik,\frac1z\right)\right.\\
\label{gtilde}
&\qquad\qquad\qquad\qquad\left.+(4z)^{ik}\Gamma(k_0-ik)\Gamma(ik)\;{}_2F_1\left(k_0-ik,\frac12-ik,1-2ik,\frac1z\right)\right],
\end{align}
where $z$ is defined as $\cosh^2(r/2)$. For $d=4$ this decomposition looks very simple:
\begin{equation}
\tilde G_k(r)=\frac{e^{k\pi}\sinh\left[k(\pi-ir)\right]}{4\pi i\sinh(k\pi)\sinh r}=\frac{e^{2k\pi-ikr}-e^{ikr}}{8\pi i\sinh(k\pi)\sinh r}.
\end{equation}
In general dimensions this decomposition may look complicated, but each term has a simple asymptotic behavior at large $r$. Specifically as $r\to\infty$, $z$ goes to $e^r/4$, both hypergeometric functions in \eqref{gtilde} goes to 1, and we have
\begin{equation}\label{gtasym}
\tilde G_k(r)\approx-\frac{1}{4\pi^{d/2}e^{k_0r}}\left[e^{2k\pi}e^{-ikr}\Gamma(k_0+ik)\Gamma(-ik)+e^{ikr}\Gamma(k_0-ik)\Gamma(ik)\right].
\end{equation}

As $k\to\pm\infty$ along the real axis, the absolute value of the products of Gamma functions in \eqref{gtasym} becomes
\begin{equation}
|\Gamma(k_0+ik)\Gamma(-ik)|=|\Gamma(k_0-ik)\Gamma(ik)|\approx2\pi |k|^{\frac{d-4}{2}}e^{-|k|\pi}.
\end{equation}
Clearly the first term in \eqref{gtasym} diverges as $k\to+\infty$, while the second term goes to zero as $k\to\pm\infty$. As we argued earlier, we treat this divergence by multiplying the first term by a factor of $e^{-2(k-ik_0)\pi}=(-1)^de^{-2k\pi}$, changing the exact form \eqref{gtilde} of $\tilde G_k(r)$ into
\begin{multline}\label{gtc}
\tilde G_k(r)=-\frac{1}{4\pi^{d/2}(4z)^{k_0}}\left[(-1)^d(4z)^{-ik}\Gamma(k_0+ik)\Gamma(-ik)\;{}_2F_1\left(k_0+ik,\frac12+ik,1+2ik,\frac1z\right)\right.\\
\left.+(4z)^{ik}\Gamma(k_0-ik)\Gamma(ik)\;{}_2F_1\left(k_0-ik,\frac12-ik,1-2ik,\frac1z\right)\right],
\end{multline}
where again $z=\cosh^2(r/2)$. This does not change the integral on $C_a$, but enables us to legally deform the contour from $C_a$ to $C$ for sufficiently negative $T+T'$. In terms of this improved $\tilde G_k(r)$ the Lorentzian correlator is
\begin{equation}\label{gfinal}
\hat G(T,T',r)=\hat G_1+\hat G_2
=\int_{-\infty}^{+\infty}\frac{dk}{2\pi}v_k(T)v_{-k}(T')G_k(r)+\int_{C}\frac{dk}{2\pi}R(k)v_{-k}(T)v_{-k}(T')\tilde G_k(r),
\end{equation}
which looks the same as \eqref{gexpl} except for the contour $C$. This is our final, exact expression for the massless correlator in any open FRW spacetime resulting from a CdL decay, valid for any $T$, $T'$, and $r$. Here $G_k(r)$ is given by \eqref{ggtilde} as
\begin{equation}\label{gkr}
G_k(r)=i^d\frac{(2\sqrt\pi)^{1-d}}{\Gamma\left(\frac{d-1}{2}\right)}\Gamma(k_0+ik)\Gamma(k_0-ik)\;{}_2F_1\left(k_0+ik,k_0-ik,\frac{d-1}{2},\cosh^2\frac r2\right).
\end{equation}

Before concluding this section, let us comment on the structure of \eqref{gfinal}. In order to calculate the Lorentzian correlator, we only need the eigenmodes $v_k(T)$ and the reflection coefficient $R(k)$. For $v_k(T)$ we solve the ``Lorentzian Schr\"odinger equation'' \eqref{schrl} with the boundary condition $v_k(T)\to e^{ikT}$ as $T\to-\infty$. For $R(k)$ we in principle need to solve the ``Euclidean Schr\"odinger equation'' \eqref{schr}. We cannot in general hope to calculate the complete Lorentzian correlator by knowing only $a(T)$ but not $a(X)$ (or vice versa); an example is provided by the thin-wall limit discussed in \cite{Freivogel:2006xu}, where the Lorentzian scale factor $a(T)=e^T$ is exactly the same as that of flat space in Milne coordinates, but the massless correlator still has a nontrivial term due to the specific vacuum chosen by the CdL geometry. From the perspective of \eqref{gfinal}, the choice of vacuum is encoded in the reflection coefficient $R(k)$ which must in general be calculated from the Euclidean scale factor $a(X)$. Usually $R(k)$ is not analytically tractable away from the thin-wall limit, but fortunately all we need to know in order to calculate the leading behavior of $\hat G_2$ is the fact that $R(k)$ always has a simple pole at $k=ik_0$, and all other poles lie below it.

\subsection{Our FRW spacetime}\label{subsec:afrw}
Let us now apply \eqref{gfinal} to our FRW model
\begin{equation}
ds_d^2=-dt^2+a(t)^2dH_{d-1}^2,
\end{equation}
where the scale factor $a(t)$ asymptotes to $ct$ at late times. Near the big bang singularity $t=0$ the scale factor $a(t)=t+\mathcal O(t^3)$ as required by the smoothness of the CdL instanton. Going to the conformal coordinates $T=\int dt/a(t)$, the scale factor $a(T)$ approaches (up to constant factors) $e^{T}$ as $T\to-\infty$, and approaches $e^{cT}$ as $T\to+\infty$. Therefore the ``Schr\"odinger potential'' $U(T)$ asymptotes to $k_0^2$ as $T\to-\infty$ and $c^2k_0^2$ as $T\to+\infty$. This means that the eigenmodes $v_k(T)$ have the following asymptotic behavior
\begin{align}
v_k(T)&\to e^{ikT}\qquad(T\to-\infty),\\
v_k(T)&\to\alpha_ke^{iT\sqrt{k^2-(c^2-1)k_0^2}}+\beta_ke^{-iT\sqrt{k^2-(c^2-1)k_0^2}}\qquad(T\to+\infty),
\end{align}
where $\alpha_k$ and $\beta_k$ are analogous to reflection and transmission coefficients. For $k>k_0\sqrt{c^2-1}$ the eigenmodes are oscillatory at large $T$; for smaller $k$ they are exponentially decaying or growing. These are pseudotachyonic modes.

The exact eigenmodes $v_k(T)$ are difficult to solve analytically except for perhaps unrealistically simple $U(T)$ such as a step potential. Fortunately, we are most interested in the equal-time massless correlator at large $T$ and large $r$, for which both $v_k(T)$ and $\tilde G_k(r)$ simplify. Specifically in the large $T$ limit $v_{-k}(T)v_{-k}(T)$ in $\hat G_2$ becomes
\begin{align}
v_{-k}(T)^2\approx\alpha_{-k}^2e^{-2iT\sqrt{k^2-(c^2-1)k_0^2}}+\beta_{-k}^2e^{2iT\sqrt{k^2-(c^2-1)k_0^2}}+2\alpha_{-k}\beta_{-k}\equiv v_-+v_++v_0,
\end{align}
where we have changed the sign of the exponent in $v_{-k}(T)$ because we choose the square root to have a branch cut between $k=\pm k_0\sqrt{c^2-1}$. In the large $r$ limit \eqref{gtc} simplifies to
\begin{equation}
\tilde G_k(r)\approx-\frac{1}{4\pi^{d/2}e^{k_0r}}\left[(-1)^de^{-ikr}\Gamma(k_0+ik)\Gamma(-ik)+e^{ikr}\Gamma(k_0-ik)\Gamma(ik)\right]\equiv\tilde G_-+\tilde G_+.
\end{equation}
Therefore the integrand of $\hat G_2$ in \eqref{gfinal} can be decomposed into six terms:
\begin{equation}
\hat G_2(T,T,r)\approx\int_C\frac{dk}{2\pi}R(k)(v_-+v_++v_0)(\tilde G_-+\tilde G_+),
\end{equation}
where for large $|k|$ we have $v_\pm\sim e^{\pm i2kT}$ and $\tilde G_\pm\sim e^{\pm ikr}/\sinh(k\pi)$ up to powers of $k$.

In the large $T$, large fixed $r$ limit that we are interested in (i.e.\ $T\gg r\gg1$) we may deform the contour from $C$ to $C_a$ for the three terms $v_+(\tilde G_-+\tilde G_+)+v_0\tilde G_+$, and deform the contour to $C_b$ for the other three: $v_-(\tilde G_-+\tilde G_+)+v_0\tilde G_-$. If we are interested in the large $r$, large fixed $T$ limit (i.e.\ $r\gg T\gg1$) instead, we simply switch the contours for $v_+\tilde G_-$ and $v_-\tilde G_+$. In both cases (or more generally, in any large $T$, $r$ limit), the leading behavior of $\hat G_2$ can be shown to come from the residue of the $v_-\tilde G_-$ term at the double pole $k=ik_0$. This can be evaluated up to numerical factors:
\begin{equation}\label{g2}
\hat G_2(T,T,r)\sim e^{2ck_0T}(2T+r),\qquad
G_2(T,T,r)=\frac{\hat G_2(T,T,r)}{a(T)^{k_0}a(T)^{k_0}}\sim 2T+r,
\end{equation}
where $G_2$ is the corresponding piece of the original correlator for $\phi$. The first term $2T$ is a pure gauge because it is linear in the coordinates. The second term $r$ is not, because it is actually the geodesic distance on $\mathbb H_{d-1}$ between the two points of the correlation function.

To study the holographic dual of our FRW model, we rewrite the metric in hyperbolic slicing
\begin{equation}
ds_d^2=-dt^2+a(t)^2(d\chi^2+\cosh^2\chi\, dH_{d-2}^2),\qquad
dH_{d-2}^2=d\tilde\chi^2+\sinh^2\tilde\chi\, d\Omega_{d-3}^2.
\end{equation}
Let us put both points of the correlator on a hypersurface of constant $\chi$, with a separation of $\Delta\tilde\chi$ in the $\tilde\chi$ direction. Their geodesic distance on $\mathbb H_{d-1}$ is therefore
\begin{equation}\label{geod}
r=\text{arccosh}(\cosh\Delta\tilde\chi\cosh^2\chi-\sinh^2\chi)\approx\Delta\tilde\chi+2\log\cosh\chi,
\end{equation}
where the approximation holds for large $\Delta\tilde\chi$ and any $\chi$. For large $\chi$ the second term on the right hand side of \eqref{geod} simply becomes $2\chi$, but let us not commit ourselves to that limit. In terms of $\chi$ and $\Delta\tilde\chi$ the leading behavior of $G_2$ from \eqref{g2} becomes
\begin{equation}
G_2(t,\chi,\Delta\tilde\chi)\sim\Delta\tilde\chi+2\log\cosh\chi.
\end{equation}
This term is present for both $c>1$ and $c=1$, and it is independent of the dimension $d$. This suggests an interpretation as the contribution from the zero mode of $\phi$ localized on the UV slice.

The leading behavior of the other term in the correlator, namely $\hat G_1$, is dominated by the contributions of the pseudotachyonic modes. The most pseudotachyonic mode is the one with $k=0$, given by
\begin{equation}
v_0(T)\sim e^{k_0T\sqrt{c^2-1}}.
\end{equation}
In the large $r$ limit $G_0(r)$ as defined in \eqref{gkr} goes like $re^{-k_0r}$, so the first term in the correlator \eqref{gfinal} becomes
\begin{equation}
\hat G_1(T,T,r)\sim re^{2k_0T\sqrt{c^2-1}-k_0 r},\quad
G_1(T,T,r)\sim re^{2k_0T(\sqrt{c^2-1}-c)-k_0 r},
\end{equation}
which we rewrite in terms of $t$, $\chi$, and $\Delta\tilde\chi$ as
\begin{equation}
G_1(t,\chi,\Delta\tilde\chi)\sim\frac{t^{(d-2)(\sqrt{1-1/c^2}-1)}}{\cosh^{d-2}\chi}e^{-(d-2)\Delta\tilde\chi/2}(\Delta\tilde\chi+2\log\cosh\chi).
\end{equation}
This term is present only for $c>1$, in which case there is a warped geometry with a consistent description in terms of a low-energy $(d-1)$-dimensional dual.

\subsection{Massive correlation functions}\label{subsec:amass}
We have hitherto focused on massless correlation functions, but it is rather straightforward to generalize it to massive correlation functions, as we outline below. This provides an alternative way of calculating the massive Green's functions for the KK modes, closed strings, and 7-7 strings in Section \ref{subsec:magf}.

Let us consider a scalar field $\phi$ with mass $m(t)$ which we allow to be time-dependent. We can use the same techniques developed in Sections \ref{subsec:aeuc} and \ref{subsec:alor} to calculate its correlation function. The only difference is that we need to add a corresponding mass term $a(X)^2m(X)^2$ to the ``Schr\"odinger potential'' $U(X)$, and similarly add $-a(T)^2m(T)^2$ to $U(T)$. This makes the potential $U(X)$ shallower, and the bound states either have larger eigenvalues (for small masses) or simply disappear (for large masses). This means that the reflection coefficient $R(k)$ no longer has a pole at $k=ik_0$. The pole is moved down to between $k=ik_0$ and $k=0$ if there is still a bound state. Therefore, the formulae for both the Euclidean correlator \eqref{gexp2} and the Lorentzian correlator \eqref{gfinal} are fully correct, as long as we define the contour $C$ to go from $k=-\infty$ to $k=+\infty$, in a way that goes just below $k=ik_0$ but always above any possible poles corresponding to bound states.

The pseudotachyonic modes disappear for sufficiently large masses due to the additional term $-a(T)^2m(T)^2$ in the ``Schr\"odinger potential'' $U(T)$. One can show that in this case the leading term in the correlation function agrees with the estimates in Section \ref{subsec:magf}.


\bibliographystyle{JHEP}
\renewcommand{\refname}{Bibliography}
\addcontentsline{toc}{section}{Bibliography}
\providecommand{\href}[2]{#2}\begingroup\raggedright

\end{document}